\documentclass[12pt,a4paper]{article}
\usepackage{graphics,graphicx}
\usepackage{amssymb,epsfig,amsmath,euscript,array}
\usepackage{cite}

\makeatletter
\@addtoreset{equation}{section}
\makeatother
\renewcommand{\theequation}{\thesection.\arabic{equation}}

\newcommand{\startappendix}{
\setcounter{section}{0}
\renewcommand{\thesection}{\Alph{section}}
\renewcommand{\theequation}{\Alph{section}.\arabic{equation}}}

\newcommand{\Appendix}[1]{
\refstepcounter{section}
\begin{flushleft}
{\Large\bf Appendix \thesection: #1}
\end{flushleft}}

\newcounter{multieqs}

\newcommand{\be}{\begin{equation}}
\newcommand{\ee}{\end{equation}}

\newcommand{\bm}[1]{\mbox{\boldmath $#1$}}

\def\bd{\begin{document}}
\def\ed{\end{document}}
\def\nn{\nonumber}
\def\bea{\begin{eqnarray}}
\def\eea{\end{eqnarray}}
\let\bm=\bibitem
\let\la=\label

\def\npb#1#2#3{Nucl. Phys. {\bf{B#1}} #3 (#2)}
\def\plb#1#2#3{Phys. Lett. {\bf{#1B}} #3 (#2)}
\def\prl#1#2#3{Phys. Rev. Lett. {\bf{#1}} #3 (#2)}
\def\prd#1#2#3{Phys. Rev. {D \bf{#1}} #3 (#2)}
\def\cmp#1#2#3{Comm. Math. Phys. {\bf{#1}} #3 (#2)}
\def\cqg#1#2#3{Class. Quantum Grav. {\bf{#1}} #3 (#2)}
\def\nppsa#1#2#3{Nucl. Phys. B (Proc. Suppl.) {\bf{#1A}}#3 (#2)}
\def\ap#1#2#3{Ann. of Phys. {\bf{#1}} #3 (#2)}
\def\ijmp#1#2#3{Int. J. Mod. Phys. {\bf{A#1}} #3 (#2)}
\def\rmp#1#2#3{Rev. Mod. Phys. {\bf{#1}} #3 (#2)}
\def\mpla#1#2#3{Mod. Phys. Lett. {\bf A#1} #3 (#2)}
\def\jhep#1#2#3{J. High Energy Phys. {\bf #1} #3 (#2)}
\def\atmp#1#2#3{Adv. Theor. Math. Phys. {\bf #1} #3 (#2)}
\newcommand{\EQ}[1]{\begin{equation} #1 \end{equation}}
\newcommand{\AL}[1]{\begin{subequations}\begin{align} #1 \end{align}\end{subequations}}
\newcommand{\SP}[1]{\begin{equation}\begin{split} #1 \end{split}\end{equation}}
\newcommand{\ALAT}[2]{\begin{subequations}\begin{alignat}{#1} #2 \end{alignat}\end{subequations}}
\def\beqa{\begin{eqnarray}}
\def\eeqa{\end{eqnarray}}
\def\beq{\begin{equation}}
\def\eeq{\end{equation}}
\def\N{{\cal N}}
\def\sst{\scriptscriptstyle}
\def\thetabar{\bar\theta}
\def\Tr{{\rm Tr}}
\def\one{\mbox{1 \kern-.59em {\rm l}}}
 \def\Nh{\hat{N}}

\def\a{\alpha}      \def\da{{\dot\alpha}}
\def\b{\beta}       \def\db{{\dot\beta}}
\def\c{\gamma}  \def\G{\Gamma}  \def\cdt{\dot\gamma}
\def\d{\delta}  \def\D{\Delta}  \def\ddt{\dot\delta}
\def\e{\epsilon}        \def\vare{\varepsilon}
\def\f{\phi}    \def\F{\Phi}    \def\vvf{\f}
\def\h{\eta}
\def\k{\kappa}
\def\l{\lambda} \def\L{\Lambda}
\def\m{\mu} \def\n{\nu}
\def\o{\omega}
\def\p{\pi} \def\P{\Pi}
\def\r{\rho}
\def\s{\sigma}  \def\S{\Sigma}
\def\t{\tau}
\def\th{\theta} \def\Th{\Theta} \def\vth{\vartheta}
\def\X{\Xeta}
\def\z{\zeta}

\def\cA{{\cal A}} \def\cB{{\cal B}} \def\cC{{\cal C}}
\def\cD{{\cal D}} \def\cE{{\cal E}} \def\cF{{\cal F}}
\def\cG{{\cal G}} \def\cH{{\cal H}} \def\cI{{\cal I}}
\def\cJ{{\cal J}} \def\cK{{\cal K}} \def\cL{{\cal L}}
\def\cM{{\cal M}} \def\cN{{\cal N}} \def\cO{{\cal O}}
\def\cP{{\cal P}} \def\cQ{{\cal Q}} \def\cR{{\cal R}}
\def\cS{{\cal S}} \def\cT{{\cal T}} \def\cU{{\cal U}}
\def\cV{{\cal V}} \def\cW{{\cal W}} \def\cX{{\cal X}}
\def\cY{{\cal Y}} \def\cZ{{\cal Z}}

\def\ua{\underline{\alpha}}
\def\ub{\underline{\phantom{\alpha}}\!\!\!\beta}
\def\uc{\underline{\phantom{\alpha}}\!\!\!\gamma}
\def\um{\underline{\mu}}
\def\ud{\underline\delta}
\def\ue{\underline\epsilon}
\def\una{\underline a}\def\unA{\underline A}
\def\unb{\underline b}\def\unB{\underline B}
\def\unc{\underline c}\def\unC{\underline C}
\def\und{\underline d}\def\unD{\underline D}
\def\une{\underline e}\def\unE{\underline E}
\def\unf{\underline{\phantom{e}}\!\!\!\! f}\def\unF{\underline F}
\def\unm{\underline m}\def\unM{\underline M}
\def\unn{\underline n}\def\unN{\underline N}
\def\unp{\underline{\phantom{a}}\!\!\! p}\def\unP{\underline P}
\def\unq{\underline{\phantom{a}}\!\!\! q}
\def\unQ{\underline{\phantom{A}}\!\!\!\! Q}
\def\unH{\underline{H}}

\def\As {{A \hspace{-6.4pt} \slash}\;}
\def\bs {{b \hspace{-6.4pt} \slash}\;}
\def\Ds {{D \hspace{-6.4pt} \slash}\;}
\def\ds {{\del \hspace{-6.4pt} \slash}\;}
\def\ss {{\s \hspace{-6.4pt} \slash}\;}
\def\ks {{ k \hspace{-6.4pt} \slash}\;}
\def\ps {{p \hspace{-6.4pt} \slash}\;}
\def\pas {{{p_1} \hspace{-6.4pt} \slash}\;}
\def\pbs {{{p_2} \hspace{-6.4pt} \slash}\;}

\def\Fh{\hat{F}}
\def\Vh{\hat{V}}
\def\Xh{\hat{X}}
\def\ah{\hat{a}}
\def\xh{\hat{x}}
\def\yh{\hat{y}}
\def\ph{\hat{p}}
\def\xih{\hat{\xi}}
\def\psit{\tilde{\psi}}
\def\Psit{\tilde{\Psi}}
\def\tht{\tilde{\th}}
\def\lt{\tilde{\lambda}}
\def\llt{\tilde{l}}
\def\At{\tilde{A}}
\def\Qt{\tilde{Q}}
\def\Rt{\tilde{R}}
\def\Nt{\tilde{N}}

\def\at{\tilde{a}}
\def\st{\tilde{s}}
\def\ft{\tilde{f}}
\def\pt{\tilde{p}}
\def\qt{\tilde{q}}
\def\vt{\tilde{v}}
\def\nt{\tilde{n}}

\def\delb{\bar{\partial}}
\def\bz{\bar{z}}
\def\bD{\bar{D}}
\def\bB{\bar{B}}

\def\bk{{\bf k}}
\def\bl{{\bf l}}
\def\bp{{\bf p}}
\def\bq{{\bf q}}
\def\br{{\bf r}}
\def\bx{{\bf x}}
\def\by{{\bf y}}
\def\bR{{\bf R}}
\def\bV{{\bf V}}

\def\d{\delta}\def\D{\Delta}\def\ddt{\dot\delta}
\def\pa{\partial} \def\del{\partial}
\def\xx{\times}
\def\uno{\mbox{1 \kern-.59em {\rm l}}}
\def\trp{^{\top}}
\def\inv{^{-1}}
\def\dag{{^{\dagger}}}
\def\pr{^{\prime}}
\def\lan{\langle}
\def\ran{\rangle}
\def\rar{\rightarrow}
\def\lar{\leftarrow}
\def\lrar{\leftrightarrow}
\newcommand{\0}{\,\!}      
\def\one{1\!\!1\,\,}
\def\im{\imath}
\def\jm{\jmath}
\newcommand{\tr}{\mbox{tr}}
\newcommand{\slsh}[1]{/ \!\!\!\! #1}
\def\vac{|0\rangle}
\def\lvac{\langle 0|}
\def\hlf{\frac{1}{2}}
\def\ove#1{\frac{1}{#1}}
\def\Box{\square}
\def\ZZ{\mathbb{Z}}
\def\CC#1{({\bf #1})}
\def\bcomment#1{}
\def\bfhat#1{{\bf \hat{#1}}}
\def\VEV#1{\left\langle #1\right\rangle}
\newcommand{\ex}[1]{{\rm e}^{#1}} \def\ii{{\rm i}}
\def\rr{{\rm r}} \def\rs{{\rm s}}\def\rv{{\rm v}}
\def\ri{{\rm i}}\def\rj{{\rm j}}
\newcommand{\lrbrk}[1]{\left(#1\right)}
\newcommand{\sfrac}[2]{{\textstyle\frac{#1}{#2}}}

\def\Li2{{\rm Li}_2}

\font\mybb=msbm10 at 12pt
\def\bb#1{\hbox{\mybb#1}}

\font\myBB=msbm10 at 18pt
\def\BB#1{\hbox{\myBB#1}}

\begin{document}

\begin{flushright}
hep-th/0407214 \\
QMUL-PH-04-06
\end{flushright}

\vspace{20pt}

\begin{center}

{\Large \bf One-Loop Gauge Theory  Amplitudes  in  } \\
\vspace{0.2cm}
{\Large \bf $\cN =4$   Super Yang-Mills
 from MHV Vertices \\}
%

%
\vspace{33pt}

{\bf Andreas Brandhuber, Bill Spence and Gabriele Travaglini}\footnote{
{\sffamily \{\tt a.brandhuber, w.j.spence,
g.travaglini\}@qmul.ac.uk }}

{\em Department of Physics\\
Queen Mary, University of
London\\
Mile End Road, London, E1 4NS\\
United Kingdom
 }

\vspace{40pt} {\bf Abstract}

\end{center}


We propose a new, twistor string theory inspired formalism to
calculate loop amplitudes in $\cN=4$ super Yang-Mills theory. In
this approach, maximal helicity violating (MHV) tree amplitudes of
$\cN=4$ super Yang-Mills are used as vertices, using an off-shell
prescription introduced by Cachazo, Svrcek and Witten, and
combined into effective diagrams that incorporate large numbers of
conventional Feynman diagrams. As an example, we apply this
formalism to the particular class of supersymmetric MHV one-loop scattering
amplitudes with an arbitrary number of external legs in  $\cN=4$
super Yang-Mills.
Remarkably, our approach naturally leads to a
representation of the amplitudes as dispersion integrals, which we
evaluate exactly. 
This yields a new, simplified form for the MHV
amplitudes, which is equivalent to the expressions
obtained previously by Bern, Dixon, Dunbar and Kosower
using the cut-constructibility approach.


\vspace{0.5cm}

\setcounter{page}{0}
\thispagestyle{empty}
\newpage

\section{Introduction}

It has been proposed recently \cite{witten} that $\cN=4$ super
Yang-Mills (SYM) has a formulation as a topological string theory
in super twistor space. This proposal, and related formulations,
have since been actively investigated (see
\cite{witten}--\!\!\cite{csw2}). The original maximal helicity
violating (MHV) tree amplitudes given in \cite{mhv, bg} take a
surprisingly simple form, and the tree amplitudes in the gauge
theory have now been understood in the context of this topological
string theory.
Inspired by the twistor approach, the
MHV amplitudes were used as vertices by Cachazo, Svrcek and Witten
(CSW) in \cite{csw}, together with a particular off-shell
prescription, in order to present a new and extremely
efficient method of
deriving more general, non-MHV tree amplitudes.
This has been further explored in
\cite{Zhu, Georgiou:2004wu, WuZhutwo, WuZhuthree,
Bena:2004ry, dk, ggk}. The CSW method of
constructing gauge theory tree amplitudes corresponds in the
twistor picture to completely disconnected instantons of degree
one, linked by twistor space propagators, whereas the original
approach of \cite{witten} used connected higher degree instantons.
The picture emerging is that there appears to be a number of
alternative ways to compute tree amplitudes, which are related by
the degeneration of higher degree curves into curves of lower
degree \cite{rsv, rv, rsv2, gmn, Bena:2004ry}.

The study of quantum amplitudes in $\cN=4$ SYM has led
to many interesting results -
see \cite{Bern92}--\!\!\cite{Bern:2004cz}
and references therein. Much of this progress
was achieved thanks to powerful unitarity-based methods
\cite{Bern:1994cg, bdk96, Bern:1996je, bm, Bern:2004cz}, whereby
properties of loop amplitudes are derived from cutting rules and
tree amplitudes.
One of the main virtues of this approach lies in the fact that
the tree amplitudes on both sides of the cut are evaluated
on-shell, and can moreover be simplified before
constructing the loop amplitude.
This is referred to as the ``cut-constructibility" approach.
The simplest application concerns
the calculation of one-loop MHV diagrams
(see \cite{dixon,Bern:1996je} for reviews).
Here one can derive explicit expressions for the
amplitudes at one loop using the known results for
the tree-level MHV amplitudes \cite{Bern:zx}.
Recent work \cite{Anastasiou:2003kj, Anastasiou:2004ki} has also
uncovered intriguing recursion relations which link two-loop
amplitudes to those at one loop, and led to speculation about the
existence of more general  cross-order relations between $L-1$ and
$L$ loop amplitudes.

So far, twistor-inspired techniques have successfully reproduced
known MHV and non-MHV tree amplitudes and inspired efficient new
techniques to calculate new tree amplitudes. One expects that this
success will carry over to perturbative calculations at the loop
level. Whilst some initial comments on loop diagrams were already
made in \cite{witten} and, more recently, the twistor space
structure of one-loop diagrams in $\cN=4$ SYM has been analysed in
\cite{csw2}, it has so far not been clear how to relate 
loop amplitudes directly to string theory in  twistor space. 
Indeed, in the presently known twistor string theories, 
conformal supergravity does not decouple from the gauge theory, 
which implies that also supergravity fields can propagate in loops 
and the twistor/gauge theory correspondence is spoiled 
at the loop level  \cite{BerkWitt2}.

However, even without the knowledge of the correct twistor string
theory description, one can investigate the calculation of loop
amplitudes, using recent results. Firstly, the recently obtained
formulae for tree amplitudes can be fed into the known field
theoretical unitarity-based methods; and secondly, one can
attempt, using twistor intuition, to find new field theory
prescriptions to compute scattering amplitudes. 
An example of the latter is the off-shell prescription of 
CSW \cite{csw}, that uses MHV amplitudes continued off-shell 
as vertices for constructing more general non-MHV amplitudes
at the tree level. It is natural to try to apply
this to the computation of loop diagrams, 
in particular to the simplest case of the supersymmetric
one-loop MHV amplitudes, where one can link to known results. 
This will be the subject of this paper.

Specifically, we will find that twistor-inspired techniques, 
when applied to one-loop diagrams, 
naturally generate dispersion integrals, 
in terms of which the amplitudes are described.
We find that our expression for the MHV scattering amplitudes is 
in  exact agreement with the result for this class of amplitudes 
obtained previously  by Bern, Dixon, Dunbar and Kosower (BDDK)
\cite{Bern:zx}  using the cut-constructibility approach.

The evaluation of the dispersion integrals proves, rather surprisingly,
to be tractable, and provides a simpler formulation of the
amplitudes in terms of a new form for the 
so-called ``two easy masses''  box function 
- see \eqref{niceonecyrilagain}.
We expect this to be an important ingredient in
further work, both in the calculation of other loop amplitudes, and in
the development of the twistor space picture.
The appearance of dispersion integrals in this
construction is rather striking. 
Does twistor string theory presage the return of the analytic 
S-matrix?

The rest of the paper is organised as follows.
In Section 2 we briefly review
the unitarity-based cut-constructibility approach used by
BDDK  in \cite{Bern:zx} in order
to derive formulae for the one-loop MHV amplitudes.
In Section 3 we explain how MHV tree amplitudes, 
continued off-shell using a prescription equivalent
to that of \cite{csw},  lift to effective vertices and can
be used to build loop diagrams which are made of collections of
MHV vertices connected by scalar propagators. Loop amplitudes
are then obtained by summing over appropriate MHV Feynman
diagrams.
In Section 4 we calculate supersymmetric MHV amplitudes at one loop 
using the procedure outlined in Section 3.
Specifically, we will combine the off-shell MHV vertices
into one-loop diagrams, and show that this approach
yields dispersion integrals%
\footnote{For a review of dispersion relations, see, for example,
\cite{bib}.} which reproduce the known results for MHV one-loop
amplitudes (expressed in terms of scalar box integrals). The formal proof
of this is presented in Section 4. Section 5 is  devoted to the
explicit calculations of the corresponding dispersion integrals, for the case
of the $n$-particle MHV scattering amplitude.
We find a new formulation of the box functions 
appearing in the amplitude, and show that this agrees
with the known expressions for the MHV amplitudes at one loop, 
using a nine dilogarithm identity proved in an Appendix.
In Section 6 we briefly comment on vanishing one-loop amplitudes. 
Finally, we present our conclusions in Section 7.


\section{One-loop MHV amplitudes from unitarity}

In this section we will briefly review the derivation of BDDK for
the one-loop MHV $\cN=4$ super Yang-Mills amplitudes from
unitarity constraints, given in \cite{Bern:zx}. We refer the
reader to this reference for more details. We suppress constant
factors connected with dimensional regularisation where they are
not essential to the discussion.

The full one-loop $n$-point MHV amplitudes in
$\cN=4$ super Yang-Mills are proportional to the tree level
amplitudes,
\beq\label{fullampl}
A_{n;1}^{\cN=4 \, {\rm MHV}} =  A^{\rm
tree}_n\, V_n^g.
\eeq
The function $V_n^g$ is given in
terms of scalar box functions $F$ by
\begin{eqnarray}
\label{Vfunction}
 V^g_{2m+1}   &=& \sum_{r=2}^{m-1} \sum_{i=1}^n
F_{n:r;i}^{2m\,e}  + \sum_{i=1}^n F_{n:i}^{1m} \ ,
\nonumber \\
V^g_{2m}   &=& \sum_{r=2}^{m-2} \sum_{i=1}^n F_{n:r;i}^{2m\,e} +
\sum_{i=1}^n F_{n:i}^{1m} + \sum_{i=1}^{n/2} F_{n:m-1;i}^{2m\,e}
\ ,
\end{eqnarray}
or, more compactly%
\footnote{In the next equation
it is understood that
$F_{n:1;i}^{2m\,e}= F_{n:i}^{1m}$ (see Eqs.~(I.5a)-(I.5b) of
\cite{Bern:zx}).
}
\cite{csw2},
\begin{eqnarray}
\label{Vfunction-bis}
 V^g_{n}   &=& \sum_{i=1}^{n} \sum_{r=1}^{[{n\over 2}]-1}
\Bigl(
1 - {1\over 2} \delta_{{n\over 2} - 1, r}
\Bigr)\,
F_{n:r;i}^{2m\,e}
\ .
\end{eqnarray}
The basic scalar box integral $I_4$ is defined by
\beq
I_4 = -i (4\pi)^{2-\epsilon}\, \int
\frac{d^{4-2\epsilon}p}{(2\pi)^{4-2\epsilon}}\,\,
\frac{1}{p^2(p-K_1)^2(p-K_1-K_2)^2(p+K_4)^2}
\ ,
 \eeq
where dimensional regularisation is used
in order to take care of infrared divergences.
The relevant integrals arising in the one-loop MHV diagrams are
related to $I_4$ for certain choices of momenta $K$, and are
denoted by $I_{4:i}^{1m}$ and $I_{4:r;i}^{2m\,e}$. These are given
in terms of the $F$ functions in \eqref{Vfunction} by
\begin{eqnarray}\label{Ifunctions}
I_{4:i}^{1m}  &=&   \frac{-2F_{n:i}^{1m}} { t^{[2]}_{i-3}
                             t^{[2]}_{i-2} } \ , \nonumber \\
I_{4:r;i}^{2m\,e}  &=&  \frac{-2F_{n:r;i}^{2m\,e}} {
              t^{[r+1]}_{i-1}  t^{[r+1]}_{i} -
              t^{[r]}_{i}  t^{[n-r-2]}_{i+r+1} } \ ,
\end{eqnarray}
with
\beq
t_i^{[r]} = (k_i+\dots+k_{i+r-1})^2
\eeq
(the $k_i$ are the external momenta). The $F$ functions depend on
the variables $t_i^{[r]}$ for certain values of $i,r$, and involve
logarithms and dilogarithms. Appendix I of \cite{Bern:zx} gives
the explicit expressions.

It is also convenient to introduce the general scalar box function
\cite{csw2}
\beqa \nonumber
 F(p,q,P,Q) & : = & -{1\over
\epsilon^2} \Bigl[ \big(- (P+p)^2\bigr)^{-\epsilon} \, +\, \big(
-(P+q)^2\bigr)^{-\epsilon} \, -\, \big( (-P)^2\bigr)^{-\epsilon}
\, -\, \big( (-Q)^2\bigr)^{-\epsilon} \Bigr]
\\ \nonumber
&+& {\rm Li}_2 \Bigl( 1 - {P^2\over (P+p)^2 } \Bigr)
\, +\,
{\rm Li}_2 \Bigl( 1 - {P^2\over (P+q)^2 } \Bigr)
\\ \nonumber
& +&
{\rm Li}_2 \Bigl( 1 - {Q^2\over (Q+q)^2 } \Bigr)
\, + \,
{\rm Li}_2 \Bigl( 1 - {Q^2\over (Q+p)^2 } \Bigr)
\\ 
&-& {\rm Li}_2 \Bigl( 1 - {P^2 Q^2\over (P+p)^2(P+q)^2 } \Bigr)
\, +\,
{1\over 2} \log^2 \Bigl( {(P+p)^2 \over (P+q)^2} \Bigr)
\ ,
\label{boxcsw2}
\eeqa
where $P+p+Q+q=0$.
Introducing the convenient variables
\beq
\label{st} s \ := \ (P+p)^2 \ , \qquad   t \ := \ (P+q)^2 \ ,
\eeq
we can  rewrite \eqref{boxcsw2} as
\beqa
\nonumber
F(s, t, P^2,
Q^2) & : = & -{1\over \epsilon^2} \Bigl[ (-s)^{-\epsilon} \, +\,
(-t)^{-\epsilon} \, -\, \big( (-P)^2\bigr)^{-\epsilon} \, -\,
\big( (-Q)^2\bigr)^{-\epsilon} \Bigr]
\\ \nonumber
&+& {\rm Li}_2 \Bigl( 1 - {P^2\over s } \Bigr) \, +\, {\rm Li}_2
\Bigl( 1 - {P^2\over t } \Bigr) +
{\rm Li}_2 \Bigl( 1 - {Q^2\over
s} \Bigr) \, + \, {\rm Li}_2 \Bigl( 1 - {Q^2\over t } \Bigr)
\\ 
&-& {\rm Li}_2 \Bigl( 1 - {P^2 Q^2\over s\, t } \Bigr)
\,
+\,
{1\over 2} \log^2 \Bigl( {s\over t} \Bigr)
\ .
\label{boxcsw22}
\eeqa
The relation to the functions $F_{n:r;i}^{2m\,e}$ is obtained by
setting $p=p_{i-1}$, $q=p_{i+r}$, and $P=p_i + \cdots +
p_{i+r-1}$. In the following sections we will make use of the
discontinuities of the function $F$ across its branch cuts. A
derivation of these discontinuities up to 
$\cO (\e^0 )$ can be found in Appendix A
and, up to and including $\cO (\e )$ terms, in Appendix B, where
these discontinuities are evaluated 
from phase space integrals.

The one-loop MHV amplitudes were constructed in  \cite{Bern:zx}
from tree diagrams using
cuts. A given cut results in singularities in the relevant
momentum channels, and from considering all possible cuts one can
construct the full set of possible singularities. From this and
unitarity one can deduce the amplitude as given in
(\ref{fullampl}).
More explicitly, consider a cut one-loop MHV diagram where the cut
separates the external momenta $k_{{m_1}}$ and $k_{{m_1}-1}$, and
$k_{{m_2}}$ and $k_{{m_2}+1}$ (i.e.~the set of external momenta
$k_{{m_1}}, k_{{m_1}+1},...,k_{{m_2}}$ lie to the left of the cut,
and the set $k_{{m_2}+1},k_{{m_2}+2},...,k_{{m_1}-1}$ lie to the
right, with momenta labelled clockwise and outgoing). This
separates the diagram into two MHV tree diagrams
connected only by two momenta $l_1$ and $l_2$
flowing across the cut, with
\beq
\label{pl}
l_1 = l_2 + P_L
\ ,
\eeq
where  $P_L = \sum_{i={m_1}}^{{m_2}}k_i$ is the sum of the
external momenta on the left of the cut. The momenta $l_1, l_2$
are taken to be null. 
It is important to note that the resulting integrals
are not equal to the corresponding Feynman integrals where $l_{1}$ and
$l_{2}$ would be left off shell; however, the discontinuities in the 
channel under consideration are identical 
and this gives enough information to 
determine the full amplitude uniquely. 

It is immediate to see that
the integrand which arises from the cut diagram
involves the function \cite{Bern:zx}
 \beq
 \label{funct-R}
 \hat{\cR} \ := \ {
\lan m_1 - 1 \, m_1 \ran \lan l_2 \, l_1\ran \over \lan m_1 -1 \,
l_1 \ran \lan -l_1 \, m_1 \ran}\ { \lan m_2  \, m_2 + 1 \ran \lan
l_1 \, l_2 \ran \over \lan m_2  \, l_2 \ran \lan -l_2  \, m_2 + 1
\ran} \ .
 \eeq
With the help of the Schouten identity we are able to rewrite
 \beqa \lan m_1 - 1 \,
m_1 \ran \lan l_2 \, l_1 \ran & =  & \lan m_1 -1\, l_1 \ran \lan
l_2 \, m_1 \ran \, + \, \lan m_1-1\,  l_2 \ran \lan m_1 \, l_1
\ran
\ ,
\\ \nonumber
\lan m_2  \, m_2 + 1 \ran \lan l_1 \, l_2 \ran &=&
 \lan m_2 \, l_2 \ran \lan l_1 \, m_2 + 1 \ran \, + \,
\lan m_2\,  l_1 \ran \lan m_2 + 1 \, l_2 \ran \ ,
\eeqa
 and hence
\beq \label{sum-R}
 \hat{\cR} = \cR( m_1 , m_2 + 1) +  \cR( m_1-1 , m_2)
-  \cR( m_1 , m_2 ) -  \cR( m_1 -1, m_2 +1) \ ,
 \eeq
where we define $\cR( i, j )$ to be the homogeneous function of
the spinors $l_1$ and $l_2$ given by
\beq
\cR (i\, j) \, := \, {\lan i \, l_2 \ran \over \lan i \, l_1
\ran} \, {\lan j \, l_1 \ran \over \lan j \, l_2 \ran} \ .
\eeq
We now rewrite  the integrand in terms of the scalar functions
appearing in the bubble, triangle and box integrals.
Firstly, we notice that%
\footnote{In the following formula we will omit
a term proportional to an $\epsilon$-tensor contracted with
four momenta, since
it vanishes upon integration.}
\beqa\label{Mai}
 \cR (i\, j) \, &= & {\lan i \, l_2 \ran \, [l_2
\, j] \, \lan j \, l_1 \ran \, [l_1 \, i]\over \lan i \, l_1 \ran
\, [l_1 \, i] \,  \lan j \, l_2 \ran\, [l_2 \, j]} \ = \ { {\rm
Tr} \left[ {1\over 2}(1 + \gamma^5) \, \hat{l}_1 \, \hat{i} \,
\hat{l}_2 \, \hat{j} \right] \over (l_1 - i)^2 (l_2 + j)^2}
\\ \nonumber
& =& { 2  \left[ (l_1 \, i) (l_2 \, j) + (l_1 \, j) (l_2 \, i) -
(l_1 \, l_2) (i\, j) \right] \over (l_1 - i)^2 (l_2 + j)^2} \ .
\eeqa
Of course, one could  equally write $\lan l_1 \, i  \ran \, [l_1
\, i]= 2(li) =  (l_1 + i)^2 =-(l_1 - i)^2 $ for example. The
correct choice of the signs in the denominator in \eqref{Mai} is
made according to the momentum flow. In \eqref{Mai} and in the
following paragraph we have written the signs which are
appropriate for $i=m_1$, $j=m_2$. The other possible cases,
corresponding to the possible arguments of the $\cR$ functions on
the right hand side of \eqref{sum-R}, are treated in a similar
manner.

The next step consists in using  \eqref{pl} to rewrite the second
and last terms in the numerator of the last expression in
\eqref{Mai}. The cancellation of bubble and triangle integrals
then takes place upon summing over the four terms in
\eqref{sum-R}. One may anticipate this cancellation by defining an
{\it effective} function $\cR^{\rm eff}(i,j)$ - this is defined to
be $ \cR (i\, j)$ minus the terms which cancel upon the summation
involved in \eqref{sum-R}. One finds that
\beq
\label{Reff}
 \cR^{\rm eff}(i,j) =
  \frac{ -2(P_L\,i)(P_L\,j) + P_L^2(i\,j) }
            {(l_1 - i)^2 (l_2 + j)^2 }
            \ ,
\eeq
and one may use this to define a function $\cR^{\rm eff}$ in an
analogous way to \eqref{sum-R}.

Now note the identity
\beq
\label{identity}
4(Pi)(Pj) - 2P^2(ij) = (P+i)^2(P+j)^2 - P^2(P+i+j)^2
\ ,
\eeq
for any momentum $P$, where $i,j$ refer to null momenta. In other
words, the left hand side of \eqref{identity} is invariant under
\beq P \to P + a i  + b j \ , \eeq where $a$ and $b$ are arbitrary
numbers. Using this, one can show that the momentum dependence
from the numerator in \eqref{Reff} precisely cancels that from the
denominator in the definition of the functions $I$ in
\eqref{Ifunctions} in each of the four cases which arise in
$\cR^{\rm eff}$. One is finally left with the result that the cut
diagram considered gives rise to the sum of four discontinuities  
(in the same channel)   of four different $F$ functions (three
when a cut leaves only two legs on one side), one for each of the cut
box diagrams which arise
. 
The knowledge of all cuts, together with the fact that due to general arguments 
\cite{Bern:zx}
this class of amplitudes is given by a sum of scalar box integrals $F$,  is sufficient
to fix the full amplitude uniquely, and it is simply given by \eqref{Vfunction}
as a sum over all scalar box functions $F$ with all coefficients equal to $1$.

\section{Off-shell derivation of the one-loop amplitudes}

The procedure summarized above for calculating the one-loop MHV
amplitude involved the study of all possible cuts. The calculation
of cuts involves an integration over the Lorentz invariant phase
space (LIPS) measure, which puts the momenta $l_1, l_2$ crossing
the cut on-shell. This fact allows one to insert the tree MHV
amplitudes directly into the integrals of the cuts, and was also
crucial in various algebraic manipulations which led to
cancellations and the simplicity of the final expression.

In \cite{csw} a prescription for taking momenta of external lines
in MHV-amplitudes off shell was given, and used to combine tree
MHV diagrams in such a way as to produce non-MHV tree amplitudes.
In this section we will define an off-shell prescription that can
be applied to one-loop diagrams (and we expect also to higher-loop
diagrams). In Sections 4 and 5 we will show that combining MHV
vertices into one-loop diagrams using this off-shell prescription
precisely yields the MHV results given earlier.

Consider an off-shell (loop) vector $L$. It can be decomposed as
\cite{Bena:2004ry,dk}
\beq
 \label{off}
 L \ = \ l
\, + \, z \eta \ ,
\eeq
where $l^2=0$, and $\eta$ is a fixed (and arbitrary) null vector,
$\eta^2=0$; $z$ is a real number.
Equation \eqref{off} determines $z$
as a function of $L$ to be
\beq
z \ = \
{L^2 \over  2 (L \eta)}
\ .
\eeq
Notice that $L^2 = z (l+\eta)^2 $; since the null vector $l$ has a non-negative
energy component and $\eta$
is null as well, it follows that $(l+\eta)^2$ has a definite sign unless the two
vectors are proportional. This implies that  the sign of $L^2$ is directly related to
the sign of $z$.%
\footnote{If $l$ is a null vector,
$l_{\a \da} = l_{\a} \tilde{l}_{\da}$,
then its energy component is $l_0 = (1/2)(l_1 \tilde{l}_{\dot{1}} +
l_2 \tilde{l}_{\dot{2}}) > 0$, as in Minkowski space
we identify $\tilde{l}_{\da} = (l_{\a})^{\ast}$.}

We can write the null vectors $l$ and $\eta$
in terms of spinors as $l_{\a \da} =
l_{\a} \llt_{\da}$,
$\eta_{\a \da} = \eta_{\a} \tilde{\eta}_{\da}$, from
which it follows that%
\footnote{We define the spinor inner products
as $\lan \l \, \m \ran := \epsilon_{\a \beta} \l^{\a} \mu^{\beta}$,
$[\lt \, \tilde{\m} ]  :=
\epsilon_{\da \dot\beta} \lt^{\da} \tilde{\mu}^{\dot{\beta}}$.}
\beqa
\label{1}
l_\a & = & {L_{\a \da} \tilde{\eta}^{\da}
\over [ \llt \, \tilde{\eta}]
}
\ ,
\\
\label{2} \llt_{\da} & = & {\eta^\a L_{\a \da} \over \lan l \, \eta
\ran} \ .
 \eeqa
These equations coincide with the CSW prescription \cite{csw} for
determining the spinor variables  $l$ and $\llt$ associated with
the off-shell (i.e.~non-null) four-vector $L$ defined in
\eqref{off}. The denominators on the right hand sides of \eqref{1}
and \eqref{2} will be irrelevant for our applications, since the
expressions we will be dealing with are homogeneous in the spinor
variables $\eta$; we will discard them.

In order to calculate loop diagrams, we need to
re-express the integration measure $d^4L$
(which appears in the expression of loop integrals)
in terms of the new variables $l$ and $z$
introduced previously. For our purposes it is useful to consider
the product of the momentum measure  and a massless scalar propagator.
After a short calculation, one finds that
\beq
\label{mmm}
 {d^4 L \over
L^2} \ = \ d\cN (l) \, {dz \over z} \ ,
\eeq
where we have introduced the Nair measure  \cite{Nair}
\beq
 d\cN (l) := \, \lan l
\, \, dl \ran \, d^2 \tilde{l} \ - \ [ \tilde{l} \, \, d
\tilde{l}] \, d^2 l \ .
\eeq
Importantly, the product of the measure factor with 
a scalar propagator $d^4 L / L^2 $ of \eqref{mmm}
is independent of the reference vector $\eta$.
The expression \eqref{mmm} will be central for our construction 
of loop diagrams.

Note that the Lorentz invariant phase space measure for
a massless particle can be expressed  precisely
in terms of the Nair measure:
\beq
\label{nairmeas}
 d^4l \, \delta^{(+)} (l^2)\ =  \
{d\cN (l)\over 4i} \ ,
\eeq
where, as before, we write the null vector $l$ as $l_{\a \da} =
l_{\a} \tilde{l}_{\da}$, and in Minkowski space we identify
$\tilde{l} = l^{\ast}$.

We conclude this section by 
summarising the strategy that we will follow 
in order  to evaluate a generic loop diagram.

The first step consists in building MHV Feynman diagrams out
of MHV vertices. 
Each spinor variable of an MHV vertex corresponding
to an internal line is taken off-shell using the prescription of CSW outlined
in the beginning of this section. 
Internal lines are then connected by scalar
off-shell propagators which connect particles of the same spin but
opposite helicity.
Note that each MHV vertex should be
multiplied by an appropriate delta function for momentum conservation.
In the next step we express all the loop integration 
momenta  as in \eqref{off}  and use the integration measure \eqref{mmm}
which already incorporates the scalar off-shell propagators.
Finally, one has to sum over all independent diagrams obtained in this
fashion for a fixed ordering of external helicity states.

In Section 4 we carry out this programme  for the particular case
of MHV amplitudes at one loop. One of the features of this procedure 
is the fact that the integration measure is naturally 
expressed  as the product of two terms:
\begin{itemize}
\item[{\bf 1.}]
a Lorentz-invariant phase space measure, and 
\item[{\bf 2.}] an  
integration over the $z$-variables 
introduced according to \eqref{off}
(one for each loop momentum).
\end{itemize}
The phase-space measure (a two-body phase space measure, 
at one loop) will be appropriately continued to 
$D=4-2\e$ dimensions in order 
to deal with potential infrared divergences.


\section{Supersymmetric MHV amplitudes at one loop from MHV vertices}

We will now write down the expression for the one-loop supersymmetric MHV
amplitude with $n$ external legs. In performing our analysis, we
will make use of the supersymmetric formulation introduced in
\cite{Nair}, which generalises the usual MHV amplitudes. In this
setup, to each particle one associates the usual commuting spinors
$\l_\a$, $\lt_{\da}$ (in terms of which the momentum of the $i$-th
particle is $p_{\a \da}^{i} = \l_{\a}^{i} \lt_{\da}^{i}$), as well
as anticommuting variables $\eta_{A}^{i}$, where $A$ is an index
of the anti-fundamental representation of $SU(4)$. The
supersymmetric amplitude can then be expanded in powers of the
various $\cN=4$ superspace coordinates $\eta_{A}^{i}$, and each
term of this expansion corresponds to a particular scattering
amplitude in $\cN=4$ SYM. A term containing $p$ powers of
$\eta_{A}^{i}$ corresponds to a scattering process where the
$i$-th particle has helicity $h_i = 1 - p/2$.

With this in mind, the  expression for a supersymmetric $n$-valent
MHV vertex is 
\beq
{\cal V} (1, \ldots , n) \ = \
i (2 \pi)^4 \delta^{(4)} \big(\sum_{i=1}^{n} \l^i \lt^i
\big)
\, \delta^{(8)} \bigl( \sum_{i=1}^{n} \l^{i} \eta^{i} \bigr)
\
\prod_{i= 1}^{n}
{1 \over \lan i \, i +1\ran}
\ ,
\eeq
where the spinors associated to off-shell legs are chosen
according to the prescription
\eqref{off} (or, equivalently,
\eqref{1} and \eqref{2}).

\begin{figure} [ht]
\label{fig1}
\vspace{.2in}
\centerline {
\includegraphics[width=4in]{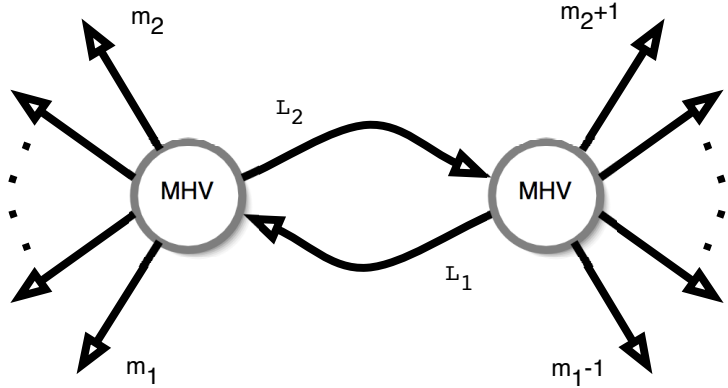}
}
\vspace{.2in}
\caption{\it One-loop MHV Feynman diagram,
computed in \eqref{mhv} using MHV amplitudes as interaction vertices,
with the CSW off-shell prescription.
The scattering amplitudes with  the desired helicities for
the external particles are then obtained by expanding the
supersymmetric scattering amplitude in powers
of the $\cN=4$ coordinates $\eta^{i}$.}
\end{figure}

We can now write down the amplitude corresponding to the
prototypical Feynman diagram we will consider, depicted in Figure
1. This is a diagram where the set of ordered external momenta
$k_1, \ldots, k_n$ is separated into two sets $k_{m_1}, \ldots ,
k_{m_2}$ (on the left), and  $k_{m_2+1}, \ldots , k_{m_1 - 1}$ (on
the right). Of course, the full one-loop MHV amplitude is 
obtained after summing over all the possible choices of these
ordered sets. We will come back to this point later, as it is
instructive to first discuss the structure of the generic one-loop
diagram we want to compute.

The expression for the Feynman diagram in Figure 1 reads
\beq
\label{mhv}
\cA \ = \ i
(2 \pi)^4 \delta^{(4)} (P_L + P_R) \ \int \! {d^4 L_1 \over L_1^2}
{d^4 L_2 \over L_2^2} \, \delta^{(4)} (L_2 - L_1 + P_L)\, 
\int\! d^8 \eta^{l_1} d^8 \eta^{l_2}
 \, \cA_L \, \cA_R \ ,
\eeq
where
$P_L = \sum_{i=m_1}^{m_2} \l^i \lt^i$ and $P_R =
 \sum_{j=m_2 + 1}^{m_1 - 1} \l^j
\lt^j$ are the sum of the momenta on the left and on the right of
the diagram, respectively, and the tree amplitudes
$\cA_L$ and $\cA_R$  are given by
\beqa
\cA_L
& = & \delta^{(8)} (\Theta_L )\
{1 \over \lan m_2 \, l_2 \ran \lan -l_1 \, m_1 \ran
\lan l_2 \, -l_1 \ran} \prod_{i= m_1}^{m_2 -
1} {1 \over \lan i \, i+1 \ran} \ ,
\\ \nonumber
\cA_R
& = &
\delta^{(8)} (\Theta_R )\
{1 \over \lan m_1-1 \, l_1 \ran
\lan -l_2 \, m_2 + 1 \ran
\lan l_1 \, -l_2 \ran}
\prod_{j= m_2+1}^{m_1 - 2}
{1 \over \lan j \, j+1 \ran}
\ .
\eeqa
Here
\beqa
\label{thetas}
(\Theta_L)_{A \a} & = &
\sum_{i\in L}\eta^{i}_A \l^{i}_{\a} \, + \,
\eta_{A}^{l_2} l_{2 \a} \, - \, \eta_{A}^{l_1} l_{1 \a}
\ ,
\\ \nonumber
(\Theta_R)_{A \a} & = &
\sum_{j\in R}\eta^{j}_A \l^{j}_{\a}\,  - \,
\eta_{A}^{l_2} l_{2 \a} \, + \, \eta_{A}^{l_1} l_{1 \a}
\ .
\eeqa
Notice that \eqref{mhv}  is  formally written
in four dimensions, but should
more correctly be analytically continued to $D=4-2 \epsilon$
dimensions, due to the presence of infrared divergences.
The effect of this is that the Lorentz invariant
phase space  measure in \eqref{LIPS} below should be taken
to be in $D=4-2 \epsilon$ dimensions. For brevity we will still
keep the four-dimensional notation.

Let us stress some important facts about \eqref{mhv}.

\begin{itemize}
\item[{\bf 1.}]
We have introduced off-shell vectors $L_1$ and
$L_2$ as in \eqref{off}, i.e.
\beq
\label{ells}
 L_{i; \a, \da} \ =
\ l_{i \a} \llt_{i \da}  \, + \,
z_i \, \eta_{\a}\tilde{\eta}_{\da} \ ,
\qquad i=1,2 \ .
 \eeq
These only appear in the measure factor
$d^4 L_1 d^4 L_2 / (L_1^2 L_2^2)$, and in the delta function.
Using \eqref{ells},
we will  rewrite the argument of the delta function as
\beq
L_2 - L_1 + P_L
= l_2 - l_1 + P_{L; z} \ ,
\eeq
where we have defined
\beq
P_{L;z} := P_L - z \eta
\ ,
\eeq
where 
\beq
z \ := \ z_1\, - \, z_2 
\ .
\eeq
Note that we use the same $\eta$ for both the momenta $L_i,
i=1,2$.

\item[{\bf 2.}]
In the remaining part of the integrand we use, as spinors
associated to the off-shell (loop) legs with momenta
$L_1$ and $L_2$, precisely the spinors
$l_{1\a}$ and $l_{2\a}$ of \eqref{ells}; this is the
essence of the CSW prescription.

\item[{\bf 3.}]
We observe that
\beqa
\label{nol}
\nonumber
&&{d^4 L_1 \over L_1^2} {d^4 L_2 \over L_2^2}
\,
\delta^{(4)} (L_2 - L_1 + P_L)
\ = \
{dz_1 \over z_1}\, { dz_2 \over z_2} \
d\cN (l_1) \, d \cN (l_2)\  \delta^{(4)} (l_2 - l_1 + P_{L;z})
\\ \cr
&&
\ \ \ \ \ \ \ \ \
\  =  \
- 4 \, {dz_1 \over z_1}{ dz_2 \over z_2}
\cdot
\Big[ d^4l_1 \, \delta^{(+)} (l_{1}^2)
\ d^4l_2 \, \delta^{(+)} (l_{2}^2)
\ \delta^{(4)} (l_2 - l_1 + P_{L;z})
\Big]
\ ,
\eeqa
where we used \eqref{mmm} and \eqref{nairmeas}.
Note that the expression
\beq
\label{LIPS}
d{\rm LIPS} (l_2 , - l_1;P_{L;z}) \ := \
d^4 l_1 \, \delta^{(+)} (l_1^2) \
d^4 l_2 \, \delta^{(+)} (l_2^2 )\
\delta^{(4)} (l_2 - l_1 + P_{L;z})
\ ,
\eeq
which appears in \eqref{nol},
is nothing but the two-particle Lorentz invariant phase
space  measure. This will be crucial in the following.
We can therefore rewrite \eqref{nol} as
\beq
\label{gcar}
{d^4 L_1 \over L_1^2} {d^4 L_2 \over L_2^2}
\,
\delta^{(4)} (L_2 - L_1 + P_L)
\ = \ -4  \, {dz_1 \over z_1}{ dz_2 \over z_2}
\,
d{\rm LIPS} (l_2 , - l_1;P_{L;z})
\ .
\eeq
\end{itemize}
To deal with infrared divergences, we will dimensionally regularise
the LIPS measure appearing  in \eqref{gcar} 
to $D= 4 - 2\epsilon$ dimensions.

Now we return to the evaluation of  \eqref{mhv}.
We start off by integrating
out the fermionic loop variables. This is easily accomplished by
first writing
\beq
\delta^{(8)} (\Theta_L) \delta^{(8)} (\Theta_R)
\ = \
 \int \! d^8 \theta d^8 \theta^{'} \
e^{ i \theta_{a}^{A} (\Theta_{L})_{A}^{a}}
\
e^{ i \theta_{a}^{' A} (\Theta_{R})_{A}^{a}}
\ ,
\eeq
where $\Theta_L$ and $\Theta_R$ are given in
\eqref{thetas}. Then
\beq
\label{jacob-f}
\int \! d^4 \eta^{l_1} d^4 \eta^{l_2} \
e^{i (\eta^{l_1}_A l_{1}^{\a} \, -\,  \eta^{l_2}_A l_{2}^{\a} )
(\theta^{\prime} - \theta)_{\a}^{A}} \ = \
\lan l_1 \, l_2 \ran^4 \,
\delta^{(8)} (\theta^{\prime} - \theta)
 \ .
\eeq
In this way, \eqref{mhv} is recast as
 \beq \label{mhv2}
  \cA \
= \ \cA_{\rm tree} \cdot \cL \ ,
 \eeq
where the tree-level amplitude $\cA_{\rm tree}$ is given by
\beq
\cA_{\rm tree} = i (2 \pi)^4
\delta^{(4)} (P_L + P_R) \ \delta^{(8)}
 \big( \sum_{i=1}^{n} \eta^{i}_A \l^{i}_{\a}\big) \,
\prod_{i= 1}^{n}
{1 \over \lan i \, i+1 \ran}
\ ,
\eeq
and the integral $\cL$ is
\beq
\label{L}
\cL \ = \
\int \! {d^4 L_1 \over L_1^2} {d^4 L_2 \over L_2^2}
\delta^{(4)} (L_2 - L_1 + P_L)
\
{ \lan m_1 - 1 \, m_1 \ran \lan l_2 \, l_1\ran
\over \lan m_1 -1 \, l_1 \ran
\lan -l_1 \, m_1 \ran}\
{ \lan m_2  \, m_2 + 1 \ran \lan l_1 \, l_2 \ran
\over \lan m_2  \, l_2 \ran
\lan -l_2  \, m_2 + 1 \ran}
\ .
\eeq
We recognize in the integrand of  \eqref{L}
the function $\hat{\cR}$ defined in \eqref{funct-R}.

We will now proceed to the evaluation of \eqref{L}.
Following steps similar to those of
\eqref{funct-R}--\eqref{Reff}, we decompose $\hat{\cR}$
as in \eqref{sum-R} and we are left with a sum of four terms.
Let us focus on the term 
$-\cR (m_1 , m_2)$
in \eqref{sum-R}; the other cases are treated analogously.
This term generates a contribution to the amplitude
which is proportional to
\beq
\label{I}
\cI = \int\! {d^4 L_1 \over L_1^2} {d^4 L_2 \over
L_2^2} \delta^{(4)} (L_2 - L_1 + P_L) \ {N(P_z) \over (l_1 -
m_1)^2 \, (l_2 + m_2)^2} \ , \eeq that is \beq \cI  =  \int\!
{dz_1 \over z_1}\, { dz_2 \over z_2} \ d\cN (l_1) \, d \cN (l_2)\
\delta^{(4)} (l_2 - l_1 + P_{L;z}) \ {N(P_z) \over (l_1 - m_1)^2
\, (l_2 + m_2)^2} \ ,
\eeq
where the numerator $N(P_z)$  is defined by
\beqa \label{num-z} N(P_z) & := &
 -2 (P_z \cdot m_1) (P_z
\cdot m_2) + P_z^2 (m_1 \cdot m_2)
\\ \nonumber \cr
&=& -2 (P_{L;z} \cdot m_1) (P_{L;z} \cdot m_2) + P_{L;z}^2 (m_1
\cdot m_2) \ ,
\eeqa
and
\beqa
P_z &:= &\sum_{i=m_1 + 1}^{m_2 - 1}p_i - z\, \eta \ ,
\\
P_{L;z}  &=& P_L - z\eta = \sum_{i=m_1}^{m_2}p_i - z\, \eta
 \ = \
P_z \, + \, m_1 \, + m_2
 \ .
\eeqa
$p_i = \l_i \lt_i$ is the momentum of the $i$-th particle.

\begin{figure} [ht]
\label{fig2}
\vspace{.2in}
\centerline {
\includegraphics[width=3in]{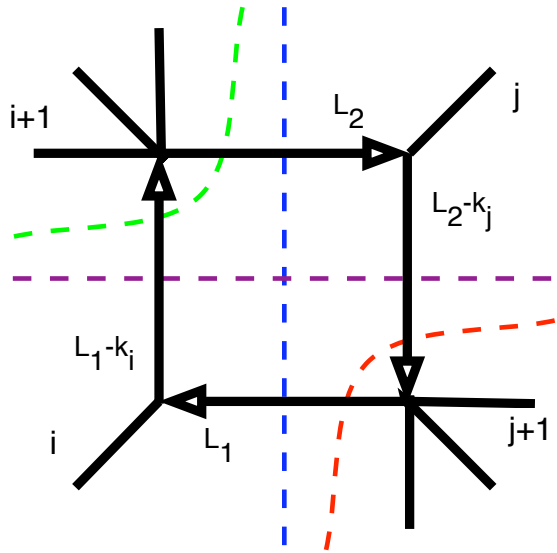}
} \vspace{.2in} \caption{\it The possible cuts of a box diagram,
corresponding to the function $F$ defined in \eqref{boxcsw22}. In
the notation of that equation, $i$ and $j$ correspond to  the
momenta $p$ and $q$, respectively; and  $P:= p_{i + 1} + \cdots +
p_{j -1}$, $Q:= p_{j+1} + \cdots + p_{i-1}$ correspond to the two
groups of  momenta on the upper left and lower right corner,
respectively. The vertical (horizontal) cuts correspond to the
$s$-channel ($t$-channel) cuts respectively, and the upper left
(lower right) corner cuts to the $P^2$-channel ($Q^2$-channel) cuts
respectively. }
\end{figure}

In the calculation reviewed in Section 2, where the MHV one-loop
amplitude is reconstructed from the cuts, one was led to
integrands involving the function defined in (\ref{Reff}). More
precisely, the numerator appearing in the corresponding box
integral is the same as that in \eqref{num-z}, {\it but evaluated at
$z=0$}.
Equivalently, our numerator \eqref{num-z} is related to that in
the cut-constructibility picture simply by the shift
\beq P_L
\longrightarrow P_{L;z} = P_L - z \eta \ .
\eeq
Next, we change variables from
$(z_1 , z_2)$ to $(z, z')$, where $z' = z_1 + z_2$, so that
\beq \label{ii} \cI \  =  2 \int\! {dz\, dz' \over (z' + z) \, (z'
- z)}\ d\cN (l_1) \, d \cN (l_2)\  \delta^{(4)} (l_2 - l_1 +
P_{L;z}) \ {N(P_{L;z}) \over (l_1 - m_1)^2 \, (l_2 + m_2)^2} \  ,
\eeq
and perform the integration over $z'$.
Equation \eqref{ii} turns into 
\beqa
\nonumber
\label{iii}
\cI \ & =&  2 (2 \pi i ) \int\!
{dz\over z}\ d\cN (l_1) \, d \cN (l_2)\ \delta^{(4)} (l_2 - l_1 +
P_{L;z}) \ {N(P_{L;z}) \over (l_1 - m_1)^2 \, (l_2 + m_2)^2}
\\
&=&
 -8 (2 \pi i ) \int\! {dz \over z}
\, d{\rm LIPS} (l_2 , - l_1; P_{L;z}) \ {N(P_{L;z}) \over (l_1 -
m_1)^2 \, (l_2 + m_2)^2}
\ .
\eeqa
In the last step we have used the two-particle Lorentz-invariant
phase space measure from \eqref{LIPS}.

Let us   briefly comment on the appearance of the LIPS measure in
\eqref{iii} and its consequences. This is the point where the
off-shell calculation presented in this paper makes contact with 
the approach
of BDDK, where one-loop amplitudes are reconstructed from the
evaluation of the cuts in the various channels. Indeed, the 
LIPS measure is precisely what is required by Cutkosky's cutting
rules \cite {Cutkosky:1960sp} to compute the discontinuity of a
Feynman diagram across the branch cuts. Which discontinuity is
evaluated is determined by the argument in the delta function
appearing in the LIPS measure; in the case of \eqref{iii}, this is
$P_{L;z}$.

Specifically, the phase space  integral appearing in \eqref{iii} is
computing a particular discontinuity of a  box diagram. The
generic box diagram, corresponding to  the function $F$ defined in
\eqref{boxcsw22} is depicted in Figure 2. The phase space integral
appearing in \eqref{iii}, which in turn was generated by the  term
$\cR (m_1 , m_2)$, corresponds to the particular (cut) box diagram
with $i=m_1$ and $j = m_2$, and where the momentum flowing in the
cut is  equal to
$\sum_{m_1}^{m_2} p_i - z \eta = - (Q + z \eta)$. It
corresponds to the lower right corner cut in Figure 2, where
however the momentum  $Q$ has been shifted to $Q_z := Q + z \eta$
(and consequently  $P \to P_z := P - z \eta$, so that momentum
conservation reads $P_z + Q_z + p + q = P + Q + p + q = 0$). Hence
we conclude that the phase space integral appearing in the last
line of \eqref{iii} is computing the $Q_z^2$-discontinuity of the
box function $F(s_z, t_z, P^2_z, Q^2_z)$, where
\beq
P_z \ := \ P
- z \eta \ , \qquad Q_z \ := \ Q + z \eta \ ,
\eeq
and
\beq s_z \ := \ (P_z + p)^2  \ ,
\qquad t_z \ := \  (P_z + q)^2 \ .
\eeq
It is now important to  remember that the full result for the one-loop
calculation of the MHV amplitudes includes a sum over various MHV
Feynman diagrams. To generate all the possible diagrams, we rotate
the external momenta in Figure 1, so that the position $m_1$ in
Figure 1 will be taken by all possible external momenta $p_1,
\ldots , p_n$. Moreover, we should also vary the number of momenta on
the left (and hence on the right) of the loop by varying the
momentum appearing in the $m_2$ position. For each diagram, we
will follow the same steps as discussed above; in particular we
will apply the Schouten identity twice, so as to generate four terms
from each diagram, as in \eqref{sum-R}.

Thus we note the following two key consequences:
\begin{itemize}
\item[{\bf 1.}]
Firstly, in this way we will produce, for each fixed box function
$F$,  exactly four phase space integrals, {\it one for each of all
the  possible cuts of the function}; and
\item[{\bf 2.}]
secondly, the  functions $F$ generated by summing over all MHV Feynman
diagrams are precisely those of the double sum of
\eqref{Vfunction-bis}, where the sum over $i$ corresponds to
different choices for the momentum in the $m_1$ position, and the
sum over $r$ corresponds to varying the number of momenta between
the legs $m_1$ and $m_2$  in Figure 1. Each function $F$ appears
precisely once (in all possible cuts).
\end{itemize}
Hence, in order to prove that  our procedure correctly generates the known
MHV one-loop amplitudes \eqref{Vfunction}, it suffices
to focus on a single function $F$.

Keeping in mind the previous considerations, we now come back to
\eqref{iii}, where the Lorentz invariant phase space integral
represents one of the four possible cuts of the  $F$ function we
wish to reproduce. These  Lorentz invariant phase space integrals
correspond to the discontinuities of the box function $F(s_z, t_z,
P^2_z, Q^2_z)$ in the four ``channels'' $s_z$, $t_z$, $P^2_z$,
$Q^2_z$. Each of the four cut-box functions is then integrated
over $z$.

Now we discuss how the final
$z$-integral of the sum of the four cuts of
 $F(s_z, t_z, P^2_z, Q^2_z)$  reproduces
the function  $F(s, t, P^2, Q^2)$, hence proving
that the procedure discussed here correctly generates
the  MHV  amplitudes at one loop.

Consider for example the $s_z$-channel, for which the variable
$P_{L;z}$ appearing in \eqref{iii} is equal to  $P + p - z \eta$.
We define
 \beq
s'\ := \ s_z \  = \ (P + p -
z\eta)^2 \ = \ s - \, 2 z \, \eta \cdot (P + p) \ ,
\eeq
where the kinematical invariant $s$ is given by $s = (P+ p)^2 $
(see \eqref{st}). It then follows that $ds' = -2 dz (\eta\, P_L)$,
and therefore \beq {dz \over z} \ = \ {ds' \over s' - s} \ . \eeq
This last step allows us to recast each of the four terms as a
dispersion integral; the dispersion integrals  reconstruct the
function $F$ from its discontinuities. The sum over Feynman
diagrams built out of MHV vertices corresponds to summing
dispersion integrals for all possible channels for each function
$F$ appearing in the sum \eqref{Vfunction}; and then summing over
all the functions appearing in \eqref{Vfunction}. In this way one
reconstructs the full MHV amplitude \eqref{fullampl} at one loop.

We will now show in more detail how the function $F (s, t, P^2,
Q^2)$ is reconstructed by summing over dispersion integrals. 
We denote by $\cF$ the
result of the $z$-integration of the sum of the four phase space
integrals (each of which is the discontinuity of $F$ in one of its
variables), that is
 \beqa
\label{disp-rel}
 \cF \ := \  \int\!
ds'\, { \Delta_{s'} F  \over s' - s} \,  \ + \ \int\! dt'\,
{\Delta_{t'} F \over t' - t} \,
\ + \ \int\! d{P^2}^{'}  \, { \Delta_{{P^2}^{'}} F  \over
{P^2}^{'} - P^2} \, \ + \ \int\! d{Q^2}^{'} { \Delta_{{Q^2}^{'}} F
\over {Q^2}^{'} - Q^2} \,
\ .
\eeqa
As is customary, for a function $f(z)$ which is analytic in the
$z$-plane except for a cut on the real axis at $z> z_0 \in
\bb{R}$, we define the discontinuity of $f(z)$ across the cut as
\beq \Delta_z\, f \ := \ f(z+i\varepsilon ) - f (z - i
\varepsilon) \ , \eeq where $\varepsilon \to 0^+$ and $z>z_0$.

Now we want to show that $\cF$ is actually proportional to the
function  $F$ itself. More precisely,  we will show that the
discontinuities of the function $\cF$  in $s$, $t$, $P^2$ and
$Q^2$ are the same as  those of the function $F$, modulo a
proportionality  coefficient. One can further argue that, in
supersymmetric theories, scattering amplitudes can be entirely
reconstructed from the knowledge of the discontinuities
\cite{Bern:1994cg} -- no ambiguities related to the presence  of
rational functions occur in this process -- to conclude that $\cF$
must actually be proportional to $F$.

To demonstrate the equality of the discontinuities, we notice that
discontinuities on the right hand side of
\eqref{disp-rel} come potentially from two sources:
\begin{itemize}
\item[{\bf 1.} ] Discontinuities due to the integral.
Those are calculated by simply remembering that
\beq
\label{dega}
{1\over x \pm i \varepsilon} = P \left( {1\over x} \right)
\, \mp \,
i \pi \, \delta (x)
 \ ,
\eeq
where $P$ stands for the principal value prescription.
\item[{\bf 2.} ]
Discontinuities in the functions $ \Delta_{s'} F$,
$ \Delta_{t'} F$, $\Delta_{{P^2}^{'}} F$,
$\Delta_{{Q^2}^{'}} F$.
\end{itemize}
However, we can immediately rule out discontinuities of type 2.
The explicit expressions of the discontinuities
 $ \Delta_{s'} F$,
$ \Delta_{t'} F$, $\Delta_{{P^2}^{'}} F$, $\Delta_{{Q^2}^{'}} F$
are listed in the Appendix, and are expressed in terms of
logarithms. The key point is that, for each integral, the
corresponding arguments of the logarithms appearing are always
positive; no discontinuities can  therefore be generated. We are
thus left with the  discontinuities of the type 1. Using
\beq
{1\over x  + i \varepsilon}\,  -\,
{1 \over  x - i \varepsilon} \ = \
-2 \pi i \, \delta (x)
\ ,
\eeq
we immediately get
\beqa
\Delta_{s} \, \cF & = &
-2\pi i \, \Delta_{s} \, F
\ ,
\\
 \nonumber
\Delta_{t} \cF &= &
-2\pi i \, \Delta_{t} \, F \ ,
\\ \nonumber
\Delta_{{P^2}} \, \cF
& = &
-2\pi i \,
\Delta_{{P^2}} \, F
\ ,
\\ \nonumber
\Delta_{{Q^2}} \, \cF
&=&
-2\pi i \, \Delta_{{Q^2}} \, F
\ .
\eeqa
We have found that the functions $\cF$ and $F$ have precisely
the same cuts in all channels ($s$, $t$, $P^2$, $Q^2$).%
\footnote{We observe that $s + t + u = P^2 + Q^2$, where $u = (p+
q)^2$.} By appealing to the cut-constructibility assumption, we
conclude that \beq \cF \ = \ -2 \pi i \, F \ . \eeq 
Finally,
notice that by picking the $\delta$-function contribution in
\eqref{dega}, we lose any dependence of the discontinuities on the
arbitrary vector $\eta$ which was used in \eqref{off} to write the
expressions for the non-null loop momenta $L_1$ and $L_2$.

In the above, we have presented a general argument 
showing that if one
combines MHV vertices with a suitable off-shell prescription, one
generates the expression \eqref{disp-rel}, giving the box function 
$\cF$ in terms a sum of dispersion integrals. 
The full $n$-particle amplitude is then given as a sum over 
box functions, as specified in Section 4.

One may wonder if the presence of infrared
singularities in the integrals affects this general argument. 
In the next section we will
present the calculation for
the $n$-gluon scattering amplitudes at one loop, showing
 how the integrals of the form \eqref{iii} give rise to
those in \eqref{disp-rel}, how the singularities in these
integrals are dealt with, and finally evaluating 
explicitly the sum of dispersion integrals appearing 
in  \eqref{disp-rel}.

\section{The $n$-particle scattering amplitude at one loop}

In this section we will consider the general MHV, one-loop, $n$-gluon
amplitude, and show that the calculation of the dispersion integrals
\eqref{disp-rel} directly yields the box function \eqref{boxcsw2}.
Summing over the MHV Feynman diagrams (Figure 1) obtained by gluing 
MHV vertices at one loop using the procedure detailed in the sections 
above then gives the sum over box functions \eqref{Vfunction-bis} which
represents the full MHV, one-loop, $n$-gluon amplitude.

As explained in the previous section, we can focus 
on a single box diagram. The goal of this section is to 
show   that the formal expression \eqref{disp-rel} 
really reproduces $F(s, t, P^2, Q^2)$.

In Appendix B we have computed the phase space integrals which 
give the discontinuities of the box diagram represented in 
Figure 2 in the channels $s$, $t$, $P^2$, $Q^2$. 
The main result is displayed in \eqref{ABM}.
It then follows that the dispersion integral in, say, 
the $s$-channel is given by
\beq \label{basicint} I(s) \ = \
\frac{1}{\epsilon} \int_0^\infty
\frac{ds'}{s-s'} (s')^{-\epsilon}(1-as')^\epsilon
\bigg[1+\epsilon^2\,{\rm Li}_2\Big(\frac{-as'}{1-as'}\Big)\bigg]
\ ,
\eeq
where
\beq
\label{a-def-part} a \ = \
\frac{P^2+Q^2-s-t}{P^2Q^2-st} \ = \
\frac{u}{P^2Q^2-st}
\ .
\eeq
A few comments are in order:
\begin{itemize}
\item[ {\bf 1.}]
In writing \eqref{basicint} we have used the form \eqref{ABM} of
the discontinuities of the box diagram, where we must also keep
the $\cO (\e )$ dilogarithm term in brackets in \eqref{ABM}.
This is because the dispersion integrals potentially give rise to
$1/ \e$  divergences,  and hence the discontinuities have to evaluated up
to order $\cO(\e)$ if the amplitude is to be calculated up to order 
$\cO(\e^{0})$.
\item[ {\bf 2.}]
In the integrand of \eqref{basicint} we have omitted  a 
numerical factor that depends on the 
dimensional regularisation parameter $\e$. 
This factor, which is  explicitly written in \eqref{ABM}, 
is irrelevant for our discussion. 
\item[ {\bf 3.}]
We have chosen the reference vector $\eta$ to be equal to $p$. We
know from the previous section that our final result will be
independent of $\eta$, at least for the cut-constructible part. However, 
what we suggest here requires a stronger gauge invariance, namely that
$\eta$ can be chosen separately for every box function, 
i.e.~$\eta$ is fixed
for the four integrals of the type \eqref{basicint} contributing to a particular box,
but can be chosen independently  
for any box.%
\footnote{A related issue was discussed in \cite{csw} to establish
independence  of tree-level amplitudes from the reference spinor 
$\eta^{\dot{\alpha}}$ needed in the definition of the MHV vertices. 
There gauge independence is only recovered after summing
over {\it all} MHV diagrams, where $\eta^{\dot{\alpha}}$ 
has to be kept fixed for all diagrams, and not only for a subset.}
Indeed it turns out that the $\eta$ dependence in the integral 
\eqref{basicint} cancels when combined with the integrals for the other
three channels. We have proven this fact for the terms, which are singular and finite
in $\epsilon$, by numerical integration for 
various arbitrary Euclidean points, i.e. $s,t,P^{2}$ and $Q^{2}$ all negative, and 
varying $\eta$ randomly.\footnote{
We thank Lance Dixon for prompting us to clarify this point and for sharing
his results on an (independent) numerical proof.}
The particular choice $\eta = p$ (or $\eta
= q$) has some advantages. One of them is that for this choice, it
turns out that the quantity $N$ introduced in \eqref{ennes}
becomes extremely simple and independent  of $z$.
Indeed, one has \beq s_z t_z \, - \,
P^2_z Q^2_z \ = \ st \, - \, P^2 Q^2 \, + \, 4z \Big[ (pq)(\eta P)
- (Pq)(\eta p) - (Pp)(\eta q) \Big] \, + \, 4 z^2 (\eta p)(\eta q)
\ , \eeq from which it follows that \beq s_z t_z \, - \, P^2_z
Q^2_z \ = \ st \, - \, P^2 Q^2 \ , \quad {\rm for} \ \eta = p \ \
{\rm or} \ \ q \ . \eeq

\end{itemize}

In the following we will choose $\eta = p$. For this choice, the
quantity $a$ defined in \eqref{a-def-gen} becomes a constant, 
and its expression  simplifies to
\eqref{a-def-part}.

The next important observation in our calculation is that 
the combination of dispersion integrals from \eqref{disp-rel} is equal
to
\beq \label{dispints}
\cF \ = \  I(s) + I(t) - I(P^2) - I(Q^2)
\ .
\eeq

Considering the combination $I(s)-I(P^2)$, expanding the
expression $(1-as')^\epsilon$ in powers of $\epsilon$, and using
Landen's identity%
\footnote{This form of Landen's identity applies for
$as' \notin (1 , \infty)$.}
\beq
\label{Landen}
\Li2 \Big( {-as' \over 1-as'} \Big) \, + \,
{1\over 2} \log^2 ( 1-as') \ = \
-\Li2 (as') \ ,
\eeq
one finds that
\beq \label{dispdiff1}
 I(s)  - I(P^2) \ = \ -\frac{s-P^2}{\epsilon}
\int_0^\infty \frac{ds'}{(s-s')(P^2-s')} (s')^{-\epsilon}
\Big[1+\epsilon\log(1-as') - \epsilon^2\,{\rm Li}_2(as')\Big]
\ .
\eeq
Now consider the three terms in the integrand above in turn. The
first may be directly integrated to yield
\beq \label{aglaia} T_1 = -\frac{1}{\epsilon^2}
\big[ \epsilon\pi \csc (\epsilon\pi)\big] \,
\Big[
(-s)^{-\epsilon} - (-P^2)^{-\epsilon} \Big]
\ .
\eeq
The second term in the integrand of \eqref{dispdiff1} gives
\beqa \label{euphrosyne} T_2 &=& -(s-P^2) \int_0^\infty
\frac{ds'}{(s-s')(P^2-s')}\, (s')^{-\epsilon}\,\log(1-as')
\\ [6pt]\nonumber
&=&
\frac{1}{\epsilon}\Big[(-a)^\epsilon(aP^2)^{-\epsilon}
\pi(as)^{-\e} \csc (\e\pi)\Big]
\, \Big[(aP^2)^\e(as)^\e\big(H(\e,aP^2)-H(\e,as)\big)
\\[6pt] \nonumber
\qquad &+& \e(as)^\e\log(1-aP^2) - \e(aP^2)^\e\log(1-as) \Big]
\ ,
\eeqa
where we define $H(\e,z) = {}_2F_1(1,\e,1+\e,z)$, with $_2F_1$ the
hypergeometric function. Now note that
\beqa \label{euphro2}
 H(\e,z)&=& {}_2F_1(1,\e,1+\e,z) = (1-z)^{-\e}\,
 {}_2F_1\Big(\e,\e,1+\e,\frac{z}{z-1}\Big)
\\ [6pt] \nonumber
&=&  (1-z)^{-\e} \Big( 1 + \e^2\,{\rm Li}_2\Big(\frac{z}{z-1}\Big)
+ \cO (\e^3) \Big)
\ . \eeqa

\noindent
Using this and expanding in powers of $\e$, one finds that
\beqa \label{euphro3}
\nonumber
 T_2 &=& [ \pi \e \csc ( \pi \e) ] \, \bigg[
{\rm Li}_2\bigg(\frac{aP^2}{aP^2-1}\bigg)
         - {\rm Li}_2\bigg(\frac{as}{as-1}\bigg)
         +\frac{1}{2}\log^2(1-aP^2) -\frac{1}{2}\log^2(1-as)
        \\ [10pt]
         &-& \log(aP^2)\log(1-aP^2) + \log(as)\log(1-as)
\biggr]
\ .
\eeqa

\noindent
Then, noting the dilogarithm identities%
\footnote{The first of these relations is again
Landen's identity \eqref{Landen}; the second one is due to Euler.}
\beqa \label{dilogids}  \nonumber
 {\rm Li}_2\Big( { x\over x-1}\Big)   \, +\,  {1\over 2}
\log^2(1-x) &=& -{\rm Li}_2(x) \ , \\ [8pt] -\Li2 (x) \, - \,
\log(x)\log(1-x) &=& \Li2(1-x) \, - \, {\pi^2\over 6} \ , \eeqa
one finds that
\beq \label{euphro4}
 T_2 \ = \  [\pi \e \csc ( \pi \e) ] \,
\Big[ \Li2(1-aP^2) - \Li2(1-as)
\Big]
\ .
\eeq

\medskip
\noindent
The third term in the integrand of \eqref{dispdiff1} gives $T_3(s)
- T_3(P^2)$, where
\beq \label{thalia}
 T_3(s) = -\e \int_0^\infty \frac{ds'}{s-s'} \,(s')^{-\e}
\, \Li2(as').
\eeq
However,
\beq \label{thalia2}
 T_3(s)\ = \  - (-s)^{-\e}\big[ \e \pi\, \csc (\e\pi)\big]
\Big[ (as)^\e{\rm LerchPhi}(as,2,\e)
 -\Li2(as)\Big] \ = \ {\cO} (\e)
\ ,
\eeq
the last equality following from the fact that ${\rm LerchPhi}
(x,2,0)=\Li2(x)$. Thus the result of this part of the integral in
\eqref{dispdiff1} vanishes as one takes $\e = 0$, and hence it 
can be dropped.

Collecting the above results \eqref{aglaia}, \eqref{euphro4},
\eqref{thalia2}, together with similar equations with $s$ and
$P^2$ replaced by $t$ and $Q^2$ respectively, we conclude that the
expression  for the box function  $F$, as obtained 
in our approach,  is
equal to (in the following expression we drop the overall,
ubiquitous factor $\pi \e \csc ( \pi \e )$)
\beqa \label{niceonecyril} \nonumber
 F (s,t,P^2, Q^2) &=&
-\frac{1}{\e^2}\Big[ (-s)^{-\e} \, + \, (-t)^{-\e} \, - \,
  (-P^2)^{-\e}\, - \,
(-Q^2)^{-\e}\Big]  \\ \nonumber  \cr
   &+& \Li2(1-aP^2)\, + \, \Li2(1-aQ^2)  \, -\,  \Li2(1-as)
\,  -\,  \Li2(1-at) \ , \\
\eeqa
where
\beq \label{aagain} a \ = \frac{P^2+Q^2-s-t}{P^2Q^2-st} 
\ = \ 
\frac{u}{P^2Q^2-st} 
\ . 
\eeq
This expression \eqref{niceonecyril} should now be compared 
with the previously known form for the function $F$
given in \eqref{boxcsw22}. 
One then concludes that \eqref{niceonecyril} and 
\eqref{boxcsw22} coincide whenever the following equality is 
satisfied:
\beqa \label{ninedilogs} \nonumber
 &&\Li2(1-aP^2) +\Li2(1-aQ^2)  - \Li2(1-as)  - \Li2(1-at)
\ =  \ \frac{1}{2}\log^2\bigg(\frac{s}{t}\bigg)
 \\  [10pt]\nonumber
 &&  +\Li2\bigg(1-\frac{P^2}{s}\bigg) + \Li2\bigg(1-\frac{P^2}{t}
\bigg) +
 \Li2\bigg(1-\frac{Q^2}{s}\bigg) + \Li2\bigg(1-\frac{Q^2}{t}\bigg)
 - \Li2\bigg(1-\frac{P^2Q^2}{st}\bigg)\
. \\ [10pt]
\eeqa
This is a remarkable identity involving nine dilogarithms.
We discuss and prove it in Appendix C.
Here we want to mention that a region in the 
space of kinematical invariants
where  the identity \eqref{ninedilogs}  holds is 
when $s$, $t$, $P^2$ and $Q^2$ are all negative, 
i.e.~the Euclidean region.
In this region  the dispersion integrals can be computed 
safely and, moreover, none of the logarithms and 
dilogarithms in \eqref{boxcsw22} have their arguments on their 
respective cuts. 
The representation \eqref{niceonecyril}
of the function $F$, which therefore coincides with  \eqref{boxcsw22}
in this region of the space of kinematical invariants, 
can then be analytically continued 
to generic values of the invariants. 
Let us also point out that  another interesting region 
where \eqref{ninedilogs} holds occurs when one of the invariants, 
say $s$, is taken to be  positive and shifted  by 
a small positive imaginary part, 
$s \to s + i \varepsilon$, $\varepsilon \to 0^+$, 
while all the other  invariants are negative.

In order to obtain the one-loop amplitude in 
a physical region, one has to perform an analytic continuation.
Usually, this continuation is simply achieved by 
the replacement 
$s \to s + i \varepsilon$, $\varepsilon \to 0^+$, where $s$ 
stands for any one of the kinematical invariants.
However, it was pointed out in 
\cite{Binoth:1999sp} that this procedure can fail if a cut is hit
by the logarithms or dilogarithms. In particular, 
the fith logarithm in \eqref{boxcsw22} is problematic \cite{Binoth:1999sp} 
and has to be amended by correction terms after analytic continuation,
see Eq.~(A.5)-(A.6) in \cite{Binoth:1999sp}.
On the other hand we have checked that 
our form of the box function \eqref{niceonecyril} 
does not suffer from this problem and, hence, 
is valid in all kinematical regions.

Summarizing, we saw in previous sections that combining MHV vertices into
one-loop diagrams yields  the dispersion integral 
representation  \eqref{disp-rel} for the box
functions. Performing the dispersion integrals explicitly, 
we have shown in this section that one obtains the
result \eqref{niceonecyril}. Using the equality
\eqref{ninedilogs}, one sees that the expression for the box
function \eqref{niceonecyril} precisely reproduces the known box
function  \eqref{boxcsw22}.
Therefore we find agreement with the known expressions 
\cite{Bern:zx}  for 
the MHV amplitudes at one loop.

We also note that the expression of the function 
$F$ given in \eqref{niceonecyril} contains only four
dilogarithms and no logarithms compared to 
\eqref{boxcsw22}, which contains five dilogarithms 
and one logarithm.

\section{Vanishing one-loop amplitudes}
In this section we make some brief remarks on the vanishing of
the simplest one-loop amplitudes, and their construction from MHV
vertices.

To begin with,  consider the one loop amplitude $\lan -\, -\ran
$ with two external gluons with negative helicity. This amplitude
violates helicity conservation and should therefore vanish.
Indeed, we can easily see this by an explicit computation
of the diagram contributing to the process built out of the
MHV vertices (continued off-shell as proposed by CSW).
It is instructive to perform the calculation
in a gauge theory with generic matter content (hence without
making reference to the Nair $\N=4$ supervertex).
Then,  summing over allowed
helicity assignments on internal legs and over allowed
particles which can circulate in the loop,
one finds that the amplitude is
proportional to a factor of
\beq 1 - n_{\rm f} + n_{\rm s} \ ,
\eeq
where $n_{\rm f}$ is the number of Weyl fermions and
$n_{\rm s}$ the number of complex scalars in the theory.
This vanishes for {\it any} supersymmetric theory.

By applying the procedure described in Section 3,
it is also easy to show that the $\cN=4$ amplitudes
$\lan - \, -\, \cdots -\ran $ vanish at one loop.
To this end, recall that for
an $L$-loop amplitude with $q$ external
gluons with negative helicity, the number $V$ of
MHV vertices which are necessary to construct it is
$V=q-1+L$.
At one loop, an amplitude with $n$ negative-helicity gluons
requires precisely $n$ MHV vertices.
A typical diagram contributing to the $n$ negative helicity gluons
amplitude is depicted in Figure 3 (for $n=5$).
It is then immediate to see that all possible cuts one can make
on the loop diagram give rise, on both sides of the cut, to tree
amplitudes which are zero. Indeed, these tree amplitudes 
must be of the form $+--\cdots -$ and hence vanish.
A similar reasoning works for the $\lan + \, -\, \cdots -\ran $
amplitude.

\begin{figure}[ht]
\label{figure-vv}
\begin{center}
\scalebox{0.47}{\includegraphics{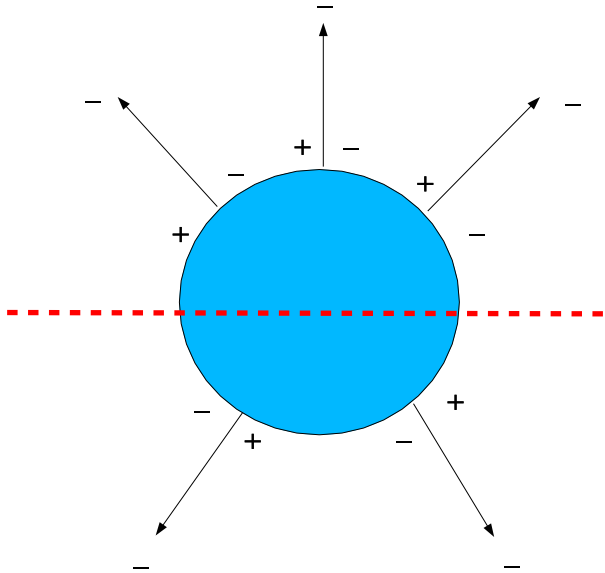}}
\end{center}
\caption{\it One of the  MHV  diagrams for the
$\lan \, -\, -\, -\, -\, -\, \ran$ process.
The tree-level diagrams on both side of the cuts
are of the type $- \cdots -\, +$, and hence vanish.
}
\end{figure}

Finally, we remark that assuming that amplitudes should
be constructible from cut diagrams and MHV tree vertices, the
amplitudes $\lan \mp \, + \cdots + \ran$ are then trivially zero
at one loop since they cannot be made from two or more MHV
vertices.

\section{Conclusions}
In this paper we have shown that supersymmetric
MHV amplitudes at one loop
can be obtained by using the tree level MHV diagrams as vertices,
complemented by the CSW prescription to take some of the external
lines off shell, when these lines form part of an internal loop.
The Feynman rules for the combination of MHV vertices into
one-loop diagrams yield a sum over terms which precisely
corresponds to the sum over terms in the dispersion relation
realization of the one-loop amplitude. This appears rather
remarkable, and is an appealing feature which one might expect to
persist in more general calculations.

In particular, one would
naturally seek to extend this work to compute next-to-maximal
helicity violating amplitudes in $\cN=4$ super Yang-Mills at one
loop using MHV vertices, which 
are only known up to  $n=6$ external  legs.  
Further applications would be to compute
more general non-MHV amplitudes for an
arbitrary number of  external gluons, as well as
the computation of higher-loop gauge theory amplitudes, and the
study of theories with less supersymmetry.

Let us also mention that our analysis provides a 
representation \eqref{niceonecyril} of the ``two easy masses'' box
function $F$ in \eqref{boxcsw22}  expressed in terms of only
four dilogarithms and with no logarithms:
 \beqa
\label{niceonecyrilagain}
 F (s,t,P^2, Q^2) &=&
-\frac{1}{\e^2}\Big[ (-s)^{-\e} \, + \, (-t)^{-\e} \, - \,
  (-P^2)^{-\e}\, - \,
(-Q^2)^{-\e}\Big]  \\ \cr
\nonumber
   &+& \Li2(1-aP^2)\, + \, \Li2(1-aQ^2)  \, -\,  \Li2(1-as)
\,  -\,  \Li2(1-at)
\ ,
\eeqa
where
\beq \label{aagainagain} a \ = \
\frac{P^2+Q^2-s-t}{P^2Q^2-st} \ = \
\frac{u}{P^2Q^2-st}
\ .
\eeq
It seems likely that this new, simpler representation will provide
clues as to the correct twistor space description of $\cN=4$ super
Yang-Mills, and might 
simplify the form of other loop amplitudes.

Finally, it is intriguing that the one-loop expression
\eqref{Vfunction-bis} for the MHV amplitudes is a sum of box
integrals - each with coefficient one. In term of the construction
with MHV vertices, this comes only after elaborating the initial
expression (by multiple applications of the Schouten identity),
and appropriately collecting cut box integrals in order to
reconstruct, from the dispersion integrals of the different cuts,
the full box function. It would be remarkable if one could find a
(string?) theory where the box diagrams represented in Figure 2 
themselves have a more fundamental and direct meaning as 
``Feynman diagrams'' emerging from a  perturbative description.


\section*{Acknowledgements}
It is a pleasure to thank Luigi Cantini,
Valya Khoze, Marco Matone and Sanjaye
Ramgoolam for helpful conversations.
GT  acknowledges the support of PPARC.

\newpage
\startappendix

\Appendix{Discontinuities of the function $F$: a first
calculation} \label{appendixnc} In this Appendix we compute the
discontinuities of the function $F(s,t,P^2,Q^2)$ defined in
\eqref{boxcsw22} considered as a function of the kinematical
invariants $s$, $t$, and the external momenta $P^2$ and $Q^2$. We
should immediately say that in this way we will miss (part of) the
$\cO (\e )$ terms in the discontinuities of the box diagrams
corresponding to the box functions (which of course does not
include all terms which vanish as $\e \to 0$). As explained in
Section 5, these $\cO (\e )$ terms are important and should be
included in the evaluation of the dispersion integrals. The
complete calculation of these terms requires the explicit
calculation of Lorentz invariant phase-space integrals, which is
presented in Appendix B. It will nevertheless be instructive to
first calculate  directly the discontinuities of the $F$ function.

Each discontinuity is computed in the following way:
all kinematical invariants are taken to be negative
except for one, say $s$, which is positive
(i.e.~in the physical region);
the discontinuity in a certain function
$f$ of the invariants is then computed as
$\Delta_s\, f =
f(s+i\varepsilon ) - f (s - i \varepsilon)$, with
$\varepsilon \to 0^+$.

The discontinuities are then given by:
\beqa
\Delta_s \, F & = &
-{1\over \epsilon^2} |s|^{-\epsilon} \, 2i \sin ( \pi \epsilon )
\, - \, 2 \pi i \Bigl[
\log {s \over -t}
\nonumber \\
&+&
\log \Bigl( 1 - {P^2 \over s} \Bigr)
+
\log  \Bigl( 1 - {Q^2 \over s} \Bigr)
-
\log  \Bigl( 1 - {P^2Q^2  \over st} \Bigr)
\Bigr]
\\ \nonumber
&=& -2\pi i \, {1\over \epsilon} |s|^{-\epsilon} \,
\, - \, 2 \pi i \log \Bigl( - {(s-P^2) (s-Q^2)
\over (st - P^2 Q^2)}
\Bigr)
\ ,
\\ \cr
\Delta_t \, F & = &
-{1\over \epsilon^2} |t|^{-\epsilon} \, 2i \sin ( \pi \epsilon )
\, -  \, 2 \pi i \Bigl[ - \log {-s \over t} +
\nonumber \\
&+& \log \Bigl( 1 - {P^2 \over t} \Bigr)
+
\log  \Bigl( 1 - {Q^2 \over t} \Bigr)
-
\log  \Bigl( 1 - {P^2Q^2  \over st} \Bigr)
\Bigr]
\\ \nonumber
&=& -2\pi i \, {1\over \epsilon} |t|^{-\epsilon}
\, - \, 2 \pi i \log \Bigl( - {(t-P^2) (t-Q^2) \over
(st - P^2 Q^2)}
\Bigr)
\ ,
\\ \cr \cr
\Delta_{P^2} \, F & = &
{1\over \epsilon^2} |P^2|^{-\epsilon} \, 2i \sin ( \pi \epsilon )
\, +  \, 2 \pi i \Bigl[ \log \Bigl( 1 - {P^2 \over s} \Bigr)
+ \log \Bigl( 1 - {P^2 \over t} \Bigr)
\nonumber \\
&-&
\log  \Bigl( 1 - {P^2Q^2  \over st} \Bigr)
\Bigr]
\\ \nonumber
&=& 2\pi i \,
{1\over \epsilon} |P^2|^{-\epsilon}
\, + \, 2 \pi i \log \Bigl(  {(s-P^2) (t-P^2) \over (st - P^2 Q^2)}
\Bigr)
\ ,
\\ \cr \cr
\Delta_{Q^2} \, F & = &
{1\over \epsilon^2} |Q^2|^{-\epsilon} \, 2i \sin ( \pi \epsilon )
\, +  \, 2 \pi i \Bigl[ \log \Bigl( 1 - {Q^2 \over s} \Bigr)
+ \log \Bigl( 1 - {Q^2 \over t} \Bigr)
\nonumber \\
&-&
\log  \Bigl( 1 - {P^2Q^2  \over st} \Bigr)
\Bigr]
\\ \nonumber
&=&
2 \pi i \,
{1\over \epsilon} |Q^2|^{-\epsilon}
\, + \, 2 \pi i \log \Bigl(  {(s-Q^2) (t-Q^2) \over (st - P^2
Q^2)} \Bigr) \ . \eeqa We notice that, for each ``channel'', the
corresponding discontinuity is perfectly well defined -- the
arguments of the various logarithms are always positive.
For example, in the $s$-channel one has $s>0$, all the other
kinematical invariants ($t$, $P^2$, $Q^2$) being negative; and
similarly for the other channels.

\Appendix{Discontinuities of the function $F$
from phase-space integrals}
Consider the phase-space integral which occurs in \eqref{iii},
\beq
\label{iiiagain}
 {\cal P} \ = \
\int\! d^{D}{\rm LIPS} (l_2, -l_1; P_{L;z}) \
{N(P_{L;z}) \over (l_1 - m_1)^2 \, (l_2 + m_2)^2}
\ ,
\eeq
where we have introduced dimensional regularisation in dimension
$D=4-2\epsilon$ \cite{'tHooft:fi} in order to  deal with infrared
divergences, with \beq \label{LIPS-dr} d^D{\rm LIPS} (l_2 , -
l_1;P_{L;z}) \ := \ d^D l_1 \, \delta^{(+)} (l_1^2) \ d^D l_2 \,
\delta^{(+)} (l_2^2 )\ \delta^{(D)} (l_2 - l_1 + P_{L;z}) \ . \eeq
The numerator $N(P_{L;z})$ is defined in \eqref{num-z} (and is a
constant with respect to the phase space integration).

The phase-space integral \eqref{iiiagain} computes 
the discontinuity of the box diagram represented in Figure 2 
in the $P_{L;z}^2$ channel. 
In the following we will use the results of
\cite{vanNeerven:1985xr,wim} to evaluate \eqref{iiiagain}.

To begin with, we observe  that the delta function in
\eqref{LIPS-dr} localizes the integral \eqref{iiiagain}
onto  vectors $l_2, l_1$ such that $l_1-l_2 = P_{L;z}$.
Lorentz transform to the centre of mass frame
of the vector $l_1-l_2$, so that
\begin{equation}
\label{comass}
l_1 \ = \    \frac{1}{2}P_{L;z} \bigl(
- 1 \, , \, -  {\bf v}
\big)\ ,
\qquad l_2 \ = \   \frac{1}{2}P_{L;z}
\bigl( -1 \, , \,  {\bf v}\big)
\ ,
\end{equation}
and write%
\footnote{We refer the reader to Appendix B of \cite{wim} for a
careful treatment of the dimensional regularisation of integrals
such as \eqref{iiiagain}.}
\begin{equation}
\label{comass2}
 {\bf v} \ = \
(\sin\theta_1\cos\theta_2\, , \, \ldots
\, , \,   \cos\theta_1)
\ .
\end{equation}
Using a further spatial rotation we write
($A,B,C$ are constants and the difference between the
four vector $m_1$ and the component
$m_1$ will be apparent from the context)
\begin{equation}  \label{comass3}
   m_1 = (m_1,0,0,m_1) \ , \qquad
   m_2 = (A,B,0,C)
\ ,
\end{equation}
with the mass-shell condition $A^2 = B^2 + C^2$.
 With this parameterisation,
\beqa
\label{carmignola}
(l_1 - m_1)^2 \, (l_2 + m_2)^2 & = &
-4\, (l_1 m_1) \, (l_2 m_2)
\\ \nonumber\cr
&\longrightarrow &
4 \, \left({P_{L; z}\over 2}\right)^2 \, m_1 \
(1 - \cos\theta_1)
(A+B\sin\theta_1\cos\theta_2+C\cos\theta_1)
\ .
\eeqa
A short calculation shows that,
after integrating over
all angular coordinates except
$\theta_1 $ and $\theta_2$,
the two-body phase space
becomes
\beq
\label{LIPS-dr2}
d^{4-2\epsilon}{\rm LIPS} (l_2, -l_1; P_{L;z}) \ = \
{
\pi^{ {1\over 2} - \epsilon} \over
4 \,  \Gamma \big({1\over 2} - \epsilon\big)
}
\,
\left| P_{L;z} \over 2 \right|^{- 2 \epsilon}
\
d\theta_1\, d\theta_2\, \,
(\sin\theta_1)^{1-2\epsilon} \,
(\sin\theta_2)^{-2\epsilon}
\ .
\eeq
Using
\eqref{LIPS-dr2} and \eqref{carmignola}, we recast
\eqref{iiiagain} as
\beq
\label{iiiagainagainagain}
{\cal P} \ =  \    \Lambda
{
\pi^{ {1\over 2} - \epsilon} \over
4 \,  \Gamma \big({1\over 2} - \epsilon\big)
}
\
\left| P_{L;z} \over 2 \right|^{- 2 \epsilon}
\
{\cal J}
\ ,
\eeq
where $\Lambda = N(P_{L;z}) / P_{L;z}^2 m_1$,
and  ${\cal J}$ is the angular integral
\beq  \label{tinkywinky}
{\cal J}\ :=\
\int_{0}^{ \pi}\!  d\theta_1
\int_{0}^{2 \pi}d\theta_2\,\!  \frac{(\sin\theta_1)^{1-2\epsilon}
(\sin\theta_2)^{-2\epsilon} }
 { (1 - \cos\theta_1)(A + C\cos\theta_1 +
B \sin\theta_1 \cos\theta_2) }
\ .
\eeq
This integral \eqref{tinkywinky}
has been evaluated in  \cite{vanNeerven:1985xr};
we borrow its result in the form of \cite{wim},
with  the result
\beq
{\cal J} \ = \
-{4\pi \over A+C} \, \left({1 \over \epsilon}\right) \,
\left( {2A \over A+C} \right)^{ \epsilon}
\,
\left[
1 \, + \, \epsilon^2\,  {\rm Li}_2 \left(
{A-C \over 2A}\right) \, + \,
{\cal O}(\epsilon^3)
\right]
\ .
\eeq
To express results in Lorentz covariant form, we note that
\begin{equation} \label{happy}
N(P_{L;z}) = -P_{L;z}^2(A+C)m_1, \quad m_1\cdot m_2 = m_1(A-C)
\ ,
\end{equation}
from which it follows that
\beqa
{1\over A + C } & = &
- {P_{L;z}^2 m_1 \over N(P_{L;z})} \ = \ - {1 \over \Lambda}
\ ,
\\
{2A\over A + C } & = &
1 \, - \, {(m_1 m_2 ) P_{L;z}^2 \over N(P_{L;z})}
\ ,
\label{banchini}
\\
\label{marcon}
{2A\over A - C } & = &
1 \, - \, { N(P_{L;z}) \over
(m_1 m_2 ) P_{L;z}^2
}
\ .
\eeqa
Putting together the above results we find that
\begin{equation}
\label{sleepy}
{\cal P} \ = \
{\pi^{ {3\over 2} - \epsilon} \over
\Gamma \big(  {1\over 2} - \epsilon \big)
}
\, \left({1 \over \epsilon}\right) \,
\left| P_{L;z}^2 \over 4 \right|^{-  \epsilon}
\
\left( {2A  \over A+ C} \right)^{ \epsilon}
\,
\left[
1 \, + \, \epsilon^2\,  {\rm Li}_2 \left(
{A-C \over 2A}\right) \, + \,
{\cal O}(\epsilon^3)
\right]
\ .
\eeq

To connect this with \eqref{disp-rel}, let $m_1=p, m_2=q$,
and note that the quantity \eqref{num-z} can be rewritten
as
\begin{equation}  \label{dopey}
2N(P_z)\  = \ P^2_z \, Q^2_z \, -\,  s_z \, t_z
\ .
\end{equation}
Using  the expressions for $2A/ (A + C) $ and
$(A-C)/ 2A$  given in \eqref{banchini}, \eqref{marcon}
(with $m_1 \to p $ and $m_2 \to q$),
we can recast \eqref{sleepy} in the form that we have used in
calculating the $n$-gluon MHV scattering amplitudes:

\begin{equation}
\label{ABM}
{\cal P} \ = \
{\pi^{ {3\over 2} - \epsilon} \over
\Gamma \big(  {1\over 2} - \epsilon \big)
}
\, \left({1 \over \epsilon}\right) \,
\left| P_{L;z}^2 \over 4 \right|^{-  \epsilon}
\
\big( 1 \, - \, a_z P_{L;z}^2  \big)^{ \epsilon}
\,
\left[
1 \, + \, \epsilon^2\,  {\rm Li}_2 \left(
{ - a_z P_{L;z}^2 \over
1 \, - \, a_z P_{L;z}^2 }
\right) \, + \,
{\cal O}(\epsilon^3)
\right]
\ ,
\eeq
where
\beq
\label{a-def-gen}
a_z\  := \ {(pq) \over N (P_{L; z})} \ = \
{u  \over 2N (P_{L; z})} \ ,
\eeq
and
\beq
u \  = \  (p+q)^2 = P^2+Q^2-s-t
\ .
\eeq

We also notice the following identities, associated to
the $s,t,P^2,Q^2$ cuts respectively:
\beqa  \label{grumpy}
-(s_z-P_z^2)(s_z-Q^2_z) &=& 2(P_{L;z}^{2(s)} p\cdot q -
N(P^{(s)}_{L;z}) )
\ ,
 \\ \nonumber
-(t_z-P_z^2)(t_z-Q^2_z) &=&
2(P_{L;z}^{2(t)} p\cdot q - N(P^{(t)}_{L;z}) )
\ ,
 \\ \nonumber
(s_z-P_z^2)(t_z-P^2_z) &=&
2(P_{L;z}^{2(P^2)} p\cdot q - N(P^{(P^2)}_{L;z}) )
\ ,
 \\ \nonumber
(s_z-Q_z^2)(t_z-Q^2_z) &=&
2(P_{L;z}^{2(Q^2)} p\cdot q - N(P^{(Q^2)}_{L;z}) )
\ ,
\eeqa
where $P_{L;z}^{(\alpha)} = P_z+p,P_z+q,P_z,Q_z$ respectively,
for labels $\alpha = s,t,P^2,Q^2$.
Note that these identities hold for any choice of the spinor
field $\eta$ which arises in the
definition of the off-shell loop momenta.
We also have
\beq
\label{ennes}
N(P^{(s)}_{L;z})\, = \,
N(P^{(t)}_{L;z})\, = \,
N(P^{(P^2)}_{L;z})\, = \,
N(P^{(Q^2)}_{L;z})\, = \,
{1\over 2} (P^2_z Q^2_z \, - \, s_z t_z)
\ .
\eeq
Using \eqref{grumpy} and \eqref{ennes},
we get the following results for the
expression of $2A / (A+C)$ in \eqref{banchini}
in the  $s$-, $t$-, $P^2$- and $Q^2$-channels:
\beqa
\label{supergrumpy}
1 - \frac{ (P^{(s)}_{L;z})^2 p \cdot q } {N(P^{(s)}_{L;z})}
& = &
-\, {(s_z-P_z^2)(s_z-Q^2_z) \over s_z t_z - P^2_z Q^2_z}
\ ,
\\ \nonumber
1 - \frac{ (P^{(t)}_{L;z})^2 p\cdot q } {N(P^{(t)}_{L;z})}
& = &
-\, {(t_z-P_z^2)(t_z-Q^2_z)\over s_z t_z - P^2_z Q^2_z}
\ ,
\\ \nonumber
1 - \frac{ (P^{(P^2)}_{L;z})^2 p\cdot q } {N(P^{(P^2)}_{L;z})}
& = &
{(s_z-P_z^2)(t_z-P^2_z)\over s_z t_z - P^2_z Q^2_z}
\ ,
\\ \nonumber
1 - \frac{ (P^{(Q^2)}_{L;z})^2 p\cdot q } {N(P^{(Q^2)}_{L;z})}
&=&
{(s_z-Q_z^2)(t_z-Q^2_z)\over s_z t_z - P^2_z Q^2_z}
\ .
\eeqa
The expressions on the right-hand sides of
\eqref{supergrumpy} are precisely those appearing in the
cuts of the box function $F$, as given in the log terms
in equations (A.1)-(A.4) of  Appendix A.
Thus we see that the
LIPS integral in \eqref{iii} generates the cuts
$\Delta F$ in the box functions.
For each cut variable $\alpha$,
we then change variables from $z$ to
$\alpha' = (P_{L;z}^{(\alpha)})^2$.
Then the sum of the four integrals
in \eqref{iii} corresponding to the four
cuts yields the expression \eqref{disp-rel} as discussed earlier.

Notice that our result for $\cP$ in \eqref{ABM} contains also an
$\cO (\e )$ term (the dilogarithm term on the right-hand side
of  \eqref{ABM}. This term could not be obtained from
the direct calculation of Appendix A, and, as explained,
is important in the calculation of dispersion integrals
of Section 5.

For reference, we write down the different values of
$((A-C)/2A)^{-1}$ (where  $((A-C)/2A)$ is the argument
of the  dilogarithm in \eqref{sleepy})
in the different channels using
\eqref{marcon} and \eqref{grumpy}:
\beqa
\label{extragrumpy}
1 - \frac{N(P^{(s)}_{L;z})   } {(P^{(s)}_{L;z})^2 p \cdot q}
& = &
-\, {(s_z-P_z^2)(s_z-Q^2_z) \over s_z \, u }
\ ,
\\ \nonumber
1 - \frac{ N(P^{(t)}_{L;z}) } {(P^{(t)}_{L;z})^2 p\cdot q }
& = &
-\, {(t_z-P_z^2)(t_z-Q^2_z)\over  t_z \, u }
\ ,
\\ \nonumber
1 - \frac{N(P^{(P^2)}_{L;z})  } {(P^{(P^2)}_{L;z})^2 p\cdot q }
& = &
{(s_z-P_z^2)(t_z-P^2_z)\over P^2_z \, u}
\ ,
\\ \nonumber
1 - \frac{N(P^{(Q^2)}_{L;z}) } { (P^{(Q^2)}_{L;z})^2 p\cdot q}
&=&
{(s_z-Q_z^2)(t_z-Q^2_z)\over Q^2_z\, u }
\ .
\eeqa
The expansion of $((A+C)/2A)^{-\epsilon}$ in \eqref{sleepy}
in powers of $\epsilon$ leads to powers of logarithms with
argument $2A / (A+C)$. These arguments are listed in
\eqref{supergrumpy} for the four possible channels.
The arguments of the dilogarithm in \eqref{sleepy}
for all possible channels are listed in \eqref{extragrumpy}.
We notice that, for each channel, whenever the argument of the
logarithm is positive, the corresponding argument of the
dilogarithm is greater than 1. In other words,
in the kinematic regime where the phase-space integral
is evaluated,  both functions (log and dilog)
are continuous functions  of their arguments.%
\footnote{In our conventions, the cut of the logarithm is on the
real negative axis; consequently, the dilogarithm has a cut on
the interval $(1, \infty)$.
}
\Appendix{Proof of the identity with nine dilogarithms}
In the calculation of the 
$n$-gluon scattering amplitude presented in Section 5 
we have encountered the  interesting 
identity \eqref{ninedilogs} involving nine dilogarithms.
The purpose of this Appendix is to give a proof of this identity.

It is instructive to present the proof for three separate cases, 
as they rely on different dilogarithm identities:%
\footnote{A list of dilogarithm identities can be found in the
relevant section of \cite{wolfram}.}
\begin{itemize}
\item[{\bf (a)}] $P^2 = Q^2 = 0$; this is the 
 case relevant for the four-gluon scattering amplitude; 
\item[{\bf (b)}] $P^2 \neq 0$, 
$  Q^2 = 0$; this is the 
 case relevant for the five-gluon scattering amplitude; and finally, 
\item[{\bf (c)}] $P^2 \neq 0$,  $Q^2 \neq 0$; this is the case 
relevant for the $n$-gluon scattering amplitudes, 
with $n \geq 6$.
\end{itemize}

We start off by addressing case {\bf (a)}. 
In this case, 
\beq
\left. a\right|_{P^2 = Q^2 = 0} \ = \ 
{1\over s }\,  + \, { 1 \over t } \ ,  
\eeq
where $a$ is defined in \eqref{aagain}. 
Using $\Li2 (1) = \pi^2 / 6$, we see that 
\eqref{ninedilogs} becomes
\beq
 -\Li2 \Big(- {  s \over t} \Big) \, - \, 
 \Li2 \Big(- {  t \over s} \Big)
\ = \ {1\over 2} \log^2 \Bigl( {s \over t} \Bigr) 
 \, + \, {\pi^2 \over 6}
\ . 
\eeq
This equation is satisfied thanks to the identity
($ z \notin (0,1)$):
\beq
\Li2 (z) \, + \,  \Li2 \Big( {1 \over z} \Big) 
\, + \, 
{1\over 2} \log^2 ( -z) \, + \, {\pi^2\over 6}
\ = \ 0
\ . 
\eeq

Next we discuss case {\bf (b)}.
In this case we make use of the relations:
\beqa
\left. 1 - as\right|_{Q^2 = 0}  & = & 1 + {u \over t} 
 \ , 
\\ \nonumber
\left. 1 - at\right|_{Q^2 = 0}  & = & 1 + {u \over s} 
\ , 
\\ \nonumber
\left. 1 - a P^2\right|_{Q^2 = 0} & = &
\Big( 1 - { P^2 \over s} \Big) 
\Big( 1 - { P^2 \over t} \Big)  
\ . 
\eeqa
Now the identity  \eqref{ninedilogs},  which we wish to prove,  turns into: 
\beqa
\label{caseb}
&& \Li2 \Big[
\Big( 1 - { P^2 \over s} \Big) 
\Big( 1 - { P^2 \over t} \Big)  \Big]
- 
\Li2
\Big( 1 - { P^2 \over s} \Big) - 
\Li2\Big( 1 - { P^2 \over t} \Big)  
 \ =   \\ \nonumber
 && \Li2 \Big( 1 + {u \over s} \Big)
+ 
\Li2 \Big( 1 + {u \over t} \Big)
+ 
{1\over 2} \log^2 \Bigl( {s \over t} \Bigr) 
\ .  
\eeqa
Eq.~\eqref{caseb} is the same as  Hill's five dilogarithm identity:
\beq
\label{Hill}
\Li2 (zw) - \Li2 (z) - \Li2 (w) - 
\Li2 \Big( { w (z-1) \over 1-w} \Big) 
-
\Li2 \Big( { z (w-1) \over 1-z} \Big) - 
{1\over 2} \log^2 \Big( { 1-w\over 1-z} \Big)\ = \ 0
\ , 
\eeq
if we make the following assignments:
\beq
z \ = \ 1 - {P^2 \over s} \ , \qquad
w \ = \ 1 - {P^2 \over t}
\ , 
\eeq
which imply 
\beq
{ w (z-1) \over 1-w}\ = \ {P^2 - t \over s} \ = \ 
1 + {u\over s} \ , 
\qquad
{ z (w-1) \over 1-z} \ = \ 
{P^2 - s \over t} \ = \ 
1 + {u\over t} \ , 
\eeq
and 
\beq
{ 1-w\over 1-z} \ = \ {s\over t}
\ . 
\eeq

Finally, we consider the general case {\bf (c)}.
The relevant identity will be Mantel's identity involving nine dilogarithms:
\beqa
\label{Mantel}
\Li2 \Bigl({ vw \over xy }\Big)  & = &
  \Li2 \Big({v\over x}\Big) 
+ \Li2 \Big({w\over x}\Big) 
+ \Li2 \Big({v\over y}\Big) 
+ \Li2 \Big({w\over y}\Big) \\ \nonumber
&+&  \Li2 (x) + \Li2 (y) - \Li2 (v) - \Li2 (w) 
+ {1\over 2} \log^2 \Big( - {x \over y } \Big) 
\ , 
\eeqa
where
\beq
\label{req}
(1-v)(1-w) \ = \ (1-x)(1-y) \ , 
\eeq
and all variables $v$, $w$, $x$, $y$, 
lie on the real axis between $0$ and $1$.

In order to use it, we notice that if we choose
\beq
\label{assig}
 x \, = \, as\ ,  
\, \qquad y \, = \, at\ , 
\qquad
 v \, = \, aP^2 \ , 
\, \qquad w \, = \, aQ^2
\ , 
\eeq
then \eqref{req} is satisfied; in particular, 
\beq
(1-v)(1-w)  \ = \  (1-x)(1-y) \  =  \ 
{(s - P^2)(s - Q^2)(t - P^2)(t - q^2) \over 
(st - P^2 Q^2)^4}
\ . 
\eeq
Now we use Mantel's identity \eqref{Mantel} 
with the assignments
\eqref{assig}, to get:
\beqa
\label{Mantel-us}
&& - \, \Li2 \Bigl({P^2 Q^2 \over st }\Big) \, + \,   
  \Li2 \Big({P^2\over s}\Big) 
\, + \,  \Li2 \Big({Q^2\over s}\Big) 
\, + \, \Li2 \Big({P^2\over t}\Big) 
\, + \, \Li2 \Big({Q^2\over t}\Big) 
\\ \nonumber
&& 
\, + \, \Li2 (as) + \Li2 (at) - \Li2 (aP^2) - \Li2 (aQ^2) 
\, + \, {1\over 2} \log^2 \Big( - {s \over t } \Big) \ = \ 0 
\ .
\eeqa
To relate the identity \eqref{Mantel-us} to 
our identity \eqref{ninedilogs}, we need a relation which 
connects $\Li2 (z)$ to $\Li2 (1-z)$. 
This is Euler's identity, 
\beq
\label{E}
\Li2(z) \  = \  
-\Li2 (1-z) \, - \, \log(z)\log(1-z)
\, + \, 
{\pi^2\over 6} \ . 
\eeq
By repeatedly using \eqref{E} in \eqref{Mantel-us}, 
we get:
\beqa \label{ninedilogs2} \nonumber
 &&
 \Li2\bigg(1-\frac{P^2Q^2}{st}\bigg)
-\Li2\bigg(1-\frac{P^2}{s}\bigg) - \Li2\bigg(1-\frac{P^2}{t}
\bigg) -
 \Li2\bigg(1-\frac{Q^2}{s}\bigg) - \Li2\bigg(1-\frac{Q^2}{t}\bigg)
\\  [10pt]\nonumber
 &&  
+ \, \Li2(1-aP^2) +\Li2(1-aQ^2)  - \Li2(1-as)  - \Li2(1-at)
\\  [10pt]\nonumber
&&
\ +\   \frac{1}{2}\log^2\bigg(- \frac{s}{t}\bigg)
\ +\  {\pi^2 \over 2} 
\ + \ 
{\cal U} \ = \ 0 \ , 
\eeqa
where 
\beqa 
&& {\cal U} \ := \ 
\nonumber
\log \bigg(\frac{P^2Q^2}{st}\bigg)
\log \bigg(1-\frac{P^2Q^2}{st}\bigg)
 - \log \bigg(\frac{P^2}{s}\bigg)\log \bigg(1-\frac{P^2}{s}\bigg)
- \log \bigg(\frac{P^2}{t}\bigg)\log \bigg(1-\frac{P^2}{t}\bigg)
\\  [10pt]
 &&  
- \log \bigg(\frac{Q^2}{s}\bigg)\log \bigg(1-\frac{Q^2}{s}\bigg)
- \log \bigg(\frac{Q^2}{t}\bigg)\log \bigg(1-\frac{Q^2}{t}\bigg)
\\  [10pt]\nonumber
 &&  
+ \log (aP^2) \log (1-aP^2)+ \log (aQ^2) \log (1-aQ^2)
- \log (as) \log ( 1- as) 
- \log (at) \log ( 1- at)  \ .  
\eeqa
Our identity \eqref{ninedilogs} is proved if 
${\cal D} = 0$, where 
\beq
\label{bol}
{\cal D}\ := \ {\cal U} \ + \   
\frac{1}{2}\log^2\bigg( \frac{s}{t}\bigg)\ + \
\frac{1}{2}\log^2\bigg(- \frac{s}{t}\bigg)
\ +\  {\pi^2 \over 2} 
\ . 
\eeq
We have checked that the right-hand side of 
\eqref{bol} indeed vanishes
whenever Mantel's identity \eqref{Mantel} is satisfied.


\end{document}